\documentclass[aps,twocolumn,pre,showpacs,amsmath,amssymb]{revtex4-1}

\usepackage{amsmath}    
\usepackage{graphicx}   
\usepackage{verbatim}   
\usepackage{color}      

\begin{document}

\title{Slow relaxation and aging kinetics for the driven lattice gas}

\author{George L. Daquila} \email{gdaquila@vt.edu} 
\affiliation{Department of Physics and Center for Stochastic Processes in 
Science and Engineering, Virginia Polytechnic Institute and State University, 
Blacksburg, VA 24061-0435}

\author{Uwe C. T\"auber} \email{tauber@vt.edu}
\affiliation{Department of Physics and Center for Stochastic Processes in 
Science and Engineering, Virginia Polytechnic Institute and State University, 
Blacksburg, VA 24061-0435}

\date{\today}

\begin{abstract}
We numerically investigate the long-time behavior of the density-density 
auto-correlation function in driven lattice gases with particle exclusion and
periodic boundary conditions in one, two, and three dimensions using precise 
Monte Carlo simulations. 
In the one-dimensional asymmetric exclusion process on a ring with half the
lattice sites occupied, we find that correlations induce extremely slow 
relaxation to the asymptotic power law decay.
We compare the crossover functions obtained from our simulations with various
analytic results in the literature, and analyze the characteristic oscillations
that occur in finite systems away from half-filling.
As expected, in three dimensions correlations are weak and consequently the 
mean-field description is adequate.
We also investigate the relaxation towards the nonequilibrium steady state in 
the two-time density-density auto-correlations, starting from strongly 
correlated initial conditions.
We obtain simple aging scaling behavior in one, two, and three dimensions, with
the expected power laws.
\end{abstract}


\maketitle

\section{Introduction}

Driven lattice gases represent perhaps the simplest models of interacting 
non-equilibrium systems that nevertheless display intriguingly complex 
features, and consequently have been carefully studied for many years now 
\cite{beatebook,hh_rev,derr1,guntbook,kirone_rev}.
Since there exists to date no general theoretical framework to adequately 
capture and classify the macroscopic properties of non-equilibrium systems, 
simple paradigmatic models such as driven lattice gases have proven useful to 
gain a thorough understanding of at least certain models driven towards 
out-of-equilibrium stationary states, characterized by non-vanishing 
macroscopic (probability, particle, or energy) currents.

Let us briefly review the basic features of driven lattice gases with particle
exclusion to provide a proper context for the present investigation.
For the one-dimensional driven lattice gas with periodic boundary conditions,
equal-time correlations have long been established to be trivial and governed
by a simple product measure \cite{derr1}.
Its temporal evolution, in contrast, displays rich features, and is not easily
accessible analytically. 
The Bethe ansatz method was used to calculate the spectral gap at half filling 
which gives the value of the dynamic exponent 
\cite{dhar,bethe_1,bethe_2,spec_osc3}. 
This approach was later generalized for arbitrary densities
\cite{spec_osc2,spec_osc1}.
The Bethe ansatz also allows for calculation of conditional time-dependent
probabilities for the lattice gas \cite{exact1,exact2}.
Yet time-dependent correlation functions have only been amenable to very 
limited general exact solutions most of which are tractable only for very small
system sizes.  
The correlation functions are governed by non-trivial scaling laws even in the
stationary state \cite{derr1}.

It is well-understood that the large-scale behavior of the driven lattice gas 
in one dimension is governed in the coarse-grained continuum limit by the noisy
Burgers equation \cite{Forster}, and is also mappable to a surface growth model
that belongs to the KPZ universality class \cite{kpz}.  
The scaling form of the density-density correlation of the driven lattice gas 
has been derived and the associated scaling exponents calculated by means of 
renormalization group techniques \cite{Forster,Jan,kpz}.  
Scaling functions have been computed by means of the mode-coupling 
approximation \cite{mc_4,mc_2,uwehwa,mc_5,cola2,mc_1,mc_3}, exact numerical 
integration \cite{cola,spohnlong,schwartz_1}, and other analytical approaches 
\cite{rg_2,rg_1,fog_1,scaling_exact_1}.

In this work, we employ extensive large-scale Monte Carlo simulations to
confirm the exactly known scaling exponents in one, two, and three dimensions.
We will also compare our numerical data to different predictions for the 
scaling functions. 
In one dimension in particular, our Monte Carlo simulations yield a remarkably
slow approach to the asymptotic scaling behavior, governed by an unusual form
of the associated scaling function.
Non-stationary properties of the driven lattice gas in the physical aging 
regime are considerably less understood.  
For KPZ surface growth, a renormalization group analysis demonstrated that the
so-called initial slip exponent can be related to the dynamical exponent, and
no additional singularities associated with the initial-time surface emerge 
\cite{krech}.  
We find that similar scaling relations hold for driven lattice gases in 
arbitrary dimensions.
In addition, for appropriate highly correlated initial conditions (that in the
growth model mapping correspond to a flat surface), we establish simple aging
scaling in the non-stationary time regime \cite{pleimbook}. 
This is in accord with studies that showed simple aging as well for both the 
KPZ surface growth model \cite{krug,A4,michelnew,yenglobal} and its linear 
counterpart, the Edwards--Wilkinson model \cite{A1,A3,A2,yenglobal}.
A few exact results were found in the non-stationary regime, for small systems 
\cite{sas_lett,sas_fluc,time_dep,sas_pape}.

The outline of this paper is as follows.  
In the following section we describe the models under investigation, comment on
our Monte Carlo simulation specifics, define the relevant observables, and 
provide the dynamic scaling forms for the density-density (and height-height)
correlation function.
In section~\ref{steady}, we present our simulation results in the 
non-equilbrium steady state for the model, focusing on the behavior in one 
dimension.  
Short time results for densities away from half filling yield charactaristic 
oscillations in the density-density correlation function.  
Next finite-size scaling features are discussed in some detail.  
We generate the scaling functions from our simulation data and compare them to
the accepted forms.  
The unusually slow relaxation of the local exponent of the density-density 
correlation function towards its asymptotic universal value is carefully 
investigated for the one-dimensional case, and contrasted with the much more 
straightforward behavior seen in higher dimensions.  
We conclude this section with a discussion of the effects of finite rather than
infinite hopping bias.  
Section~\ref{aging} contains our simulation results for non-stationary 
properties of the driven lattice gas, starting from a correlated initial state,
in one, two and three dimensions.  
We cast our results in the standard simple aging scaling form and compute the 
aging exponents by means of a simple scaling argument, which is 
{\em a-posteriori} confirmed by our simulations.

\section{The driven lattice gas with exclusion}

\subsection{Model description}

We study driven lattice gases with site exclusion on a $d$-dimensional square
lattice of size $L^d$ with periodic boundary conditions, with a total number of
$N$ particles \cite{beatebook,derr1,guntbook}.   
The exclusion constraint limits the site occupation number at a position $x$ on
the lattice to the values $n(x) \in \{0,1\}$.
The particles may hop to nearest-neighbor lattice sites only if these are 
empty.  
In one dimension, this model is referred to as the periodic asymmetric simple
exclusion process (ASEP).
We denote the probabilities of hopping ``forward'' and ``backward'' as $p$ and
$q = 1 - p$, respectively.  
For $p = q$, no net particle current arises; the system is then in equilibrium.
For $p = 1$ and hence $q = 0$, only unidirectional motion (to the right) is 
possible, and the model reduces to the totally asymmetric simple exclusion 
process (TASEP). 
The generalization of the driven lattice gas model to higher dimensions $d$ is 
straightforward. 
After a single specific drive direction is selected, we define the hopping
probabilities (to empty sites) along this direction to be $p_\parallel$ and 
$q_\parallel = 1 - p_\parallel$, while there is no hopping bias in the $d-1$ 
transverse dimensions, $p_\perp = q_\perp$ (symmetric exclusion process, SEP).

In one dimension, this lattice gas model can be mapped to a surface growth 
problem in the KPZ universality class \cite{kpz}.  
To this end, one defines local Ising spin variables $\sigma(x) = 1 - 2 n(x)$, 
where $n(x)$ is the occupation number.  
The height variable at site $x$ is then given by the accumulated magnetization
from the (arbitrary) reference point at site $1$ up to $x$:
\begin{equation}
  \label{mapp}
  h(x) = \displaystyle \sum_{y=1}^x \sigma(y) 
  = x - 2 \displaystyle \sum_{y=1}^x n(y) .
\end{equation}
Particle exclusion restricts the local slope of the surface height profile to 
$\pm 1$.  
This mapping allows us to directly relate observables such as correlation 
functions calculated for the surface growth model to those for our driven 
lattice gas model.

\subsection{Monte Carlo simulations}

We employ both standard Metropolis and continuous-time Monte Carlo algorithms 
to study the driven lattice gas.
The standard Monte Carlo algorithm \cite{New} starts with all particles placed 
at random positions.  
A particle is randomly selected, and then one of the $2 d$ directions for a 
potential nearest-neighbor hop is randomly chosen.  
A hop has a predetermined probability $R_i \in [0,1]$ ($i = 1, \ldots, 2 d$,
$R_i$ is either $p_\parallel$, $q_\parallel$, $p_\perp$, or $q_\perp$).  
A random number $r \in [0,1]$ is picked; if $r < R_i$ then hop $i$ is 
performed.  
Particle exclusion is realized in this simulation by setting $R_i = 0$ for any
hop that attempts to move a particle into an occupied site.  
One Monte Carlo time step has passed once $N$ hopping attempts have been made.
On average the algorithm thus attempts to hop each particle once per Monte 
Carlo time step.  
In dimensions $d \geq 2$, the hopping probabilities are asymmetric parallel to 
the drive (if not mentioned otherwise we use $p_\parallel = 1$, 
$q_\parallel = 0$), and symmetric in the transverse directions 
$p_\perp = q_\perp = 1/2$.  
With the periodic boundary conditions, this prescription generates a 
(fluctuating) particle current along the drive direction.

For the one-dimensional system we employ the continuous-time Monte Carlo 
algorithm \cite{Borz,New}, which is initialized the same way as the standard 
Monte Carlo process.  
The crucial feature of this algorithm is that hops are never rejected; they are
simply performed at the correct rates.  
The probability for each possible hop $i$ from a given configuration $C$ of the
system is $R_i$.  
An active list is kept of all possible hops with $R_i \not= 0$, and the next 
hop is chosen by sampling the list.  
Time is incremented by $\Delta t = 1/\sum_i R_i$.  
For the driven lattice gas with exclusion, each site has an equal probability 
of being occupied, equal to the filling fraction or mean density 
$\rho = N / L$; e.g., for a half-filled system the probability is $\rho = 1/2$.
This algorithm eliminates all attempted hops that would be rejected due to 
exclusion.  
This is realized in the simulation as follows for the ASEP.  
Two lists are maintained,  $L_r$, $L_l$, one for the hops to the right and left
with probabilities $p$ and $q$, normalized by $p + q = 1$.  
A random number $r \in [0,1]$ is generated.  
If $r < p$, a hop to the right is performed, if $r > p$ we chose a hop to the
left.  
Note that we may only proceed in this manner because the list lengths $N_h$ of 
possible left and right hops are equal. 
A random integer $n \in [1,N_h]$ is selected, and we correspondingly update the
$n$th particle in the $L_r$ or $L_l$ list.  
The time counter is increased by $\Delta t$ which is calculated using the rates
prior to the hop being performed.  
This continuous-time algorithm yields an efficiency enhancement of 
approximately a factor of $2$, because only hops which can be performed are 
chosen, and consequently allows us to increase the effective simulation time by
that factor also.
We note that in the following, all distances and times will be measured in 
units of the lattice spacing and Monte Carlo Steps (MCS), respectively.

\subsection{Quantities of interest}

Our starting point is the two-time density-density (connected) correlation 
function (cumulant), defined as
\begin{equation}
  S(\vec x',\vec x,t',t) = \langle n(\vec x,t) n(\vec x',t') \rangle 
  - \langle n(\vec x',t') \rangle \langle n(\vec x,t) \rangle ,
\end{equation}
where $\langle \cdot \rangle$ denotes an ensemble average over (many) 
stochastic realizations.  
In the stationary state, a periodic system becomes translationally invariant in
both space and time.  
Therefore the correlation function depends only on coordinate and time 
differences.  
Defining the average density $\rho = \langle n(\vec x,t) \rangle = N/L$, we 
then obtain
\begin{equation}
  S(\vec x'-\vec x,t'-t) = \langle n(\vec x,t) n(\vec x',t') \rangle - \rho^2 .
\end{equation}

For the driven lattice gas (ASEP) in the stationary state, it is well known 
that each possible particle configuration $C$ occurs with equal probability 
$P^*(C)$, namely the inverse of the number $N(C) = {L \choose N}$ of possible 
states
\begin{equation}
  \label{probC}
  P^*(C) = \frac{1}{N(C)} = \frac{N!(L-N)!}{L!} \ .
\end{equation}
This independent-particle result remarkably remains valid even in the presence
of on-site particle exclusion \cite{meak}. 
However, if nearest-neighbor interactions were added, the stationary 
probabilities would depend on the particles' relative positions.

>From the simple product measure (\ref{probC}), we can readily obtain the 
equal-time stationary density-density correlation function
$S(\vec x' - \vec x \not = 0, t' - t = 0)$.  
The probability to find a particle at a given site $\vec x$ is just the average
density $\rho = N/L$, and given that site $\vec x$ is occupied the density at a
different site $\vec x' \not= \vec x$ is $(N-1)/(L-1)$, which yields
$\langle n(\vec x,t) n(\vec x',t) \rangle = \sum_C P^*(C) n(\vec x) n(\vec x')
= \sum_{C'} P^*(C') N (N-1) / L (L-1)$.
The sum now extends over configurations $C'$, which encompass all possible 
states of $N-2$ particles in the remaining $L-2$ sites, whose stationary 
probability $P^*(C')$ is again given by Eq.~(\ref{probC}).
But the summation gives us a multiplicative factor $N(C') = 1 / P^*(C')$, which
is the number of possible configurations of the remaining $N-2$ particles 
placed in the remaining $L-2$ sites. 
Hence we arrive at the simple result 
$\langle n(\vec x,t) n(\vec x',t) \rangle = N (N-1) / L (L-1)$, and the 
equal-time correlation function becomes
\begin{equation}  
  \label{etcorr}
  S(\vec x'-\vec x \not= 0,0) = \frac{N}{L} \frac{N-1}{L-1} 
  - \left( \frac{N}{L} \right)^2 .
\end{equation}
Taking the thermodynamic limit $N,L \to \infty$ while keeping $\rho = N/L$ 
constant, the correlation function (\ref{etcorr}) vanishes for 
$\vec x \not= \vec x'$.  
However, in a finite system, Eq.~(\ref{etcorr}) yields a negative value
\begin{equation}
  \label{smin}
  S_{\rm min} = \rho \frac{(\rho L - 1)}{L - 1} - \rho^2 
  = - \frac{\rho (1 - \rho)}{L - 1} ,
\end{equation}
reflecting the particle anti-correlations induced by site exclusion.

A more interesting quantity to study is the two-time (connected) 
auto-correlation function $S(\vec x,\vec x,t',t)$, which in the stationary 
state can be written as 
$S(0,t) = \langle n(\vec x,0) n(\vec x,t) \rangle - \rho^2$.

In a finite one-dimensional system with periodic boundary conditions, i.e., a
ring with $L$ sites, another notable quantity is the velocity of a single
``tagged'' particle \cite{derr1}
\begin{equation} 
  \label{vtag}
  v_t = \frac{L-N}{L-1} \ .
\end{equation}
We have checked and confirmed this tagged particle velocity in our Monte Carlo
simulations.
Note that $v_t = 1 - \rho$ in the thermodynamic limit. 
Since our numerical data are obtained in finite systems, the typical return 
time $t_r = L / v_t$ for a specific particle to traverse the entire ring and 
come back to its original site posits an upper limit to the argument $t < t_r$
for a meaningful analysis of the time-dependent auto-correlation function 
$S(0,t)$.
We remark that the tagged particle velocity has to be carefully distinguished 
from the propagation speed of a collective density fluctuation \cite{guntbook}
\begin{equation}
\label{vcoll}
  v_c = 1 - 2 \rho ,
\end{equation}
which vanishes only at half-filling $\rho = 1/2$.

\subsection{Dynamic scaling}

In the scaling limit, the steady-state density-density correlations for the 
driven lattice gas with exclusion in $d$ dimensions are described by a general 
homogeneous function of the following form \cite{beatebook}
\begin{equation}
  \label{xtgen}
  S(\vec{x}_\perp,x_\parallel,t) = b^{d+\Delta} 
  S\Bigl( b \vec{x}_\perp, b^{1+\Delta} x_\parallel, b^z t \Bigr) ,
\end{equation}
where $b$ represents an arbitrary real scale factor, and the subscripts $\perp$
and $\parallel$ denote transverse and parallel directions with respect to the 
drive.  
The exponents $\Delta$ and $z$ are the anisotropy and (transverse) dynamical 
exponents, respectively.  
A renormalization group analysis \cite{Jan} yields the upper critical dimension
$d_c = 2$, and the exact exponent values $z=2$ and $\Delta = (2-d)/3$ for 
$d < d_c$.  
For $d > d_c$ the exponents are given by mean-field theory, i.e., $z = 2$ and
$\Delta = 0$, with logarithmic corrections at $d_c = 2$.  
Setting $b = t^{-1/z}$, Eq.~(\ref{xtgen}) becomes
\begin{equation}
  \label{scaledt}
  S(\vec{x}_\perp,x_\parallel,t) = t^{- (d + \Delta) / z} 
  f_d\Bigl( t^{-1/z}\vec{x}_\perp, t^{-1/z_\parallel} x_\parallel \Bigr) , 
\end{equation}
where $z_\parallel = z / (1 + \Delta)$ is the dynamical exponent along the 
drive direction.
Taking $\vec{x}_\perp,x_\parallel \to 0$, we obtain for the density
auto-correlation function in the scaling limit:
\begin{equation}
  \label{zeta}
  S(0,0,t) \sim t^{- \zeta} , \ 
  \zeta = \frac{d + \Delta}{z} = \frac{d + \Delta}{(1 + \Delta) z_\parallel}\ .
\end{equation}
In order to carefully characterize the approach to the asymptotic regime in our
Monte Carlo simulations, we shall utilize a corresponding local (effective) 
exponent defined via
\begin{equation}
  \label{eq:loc}
  - \zeta_{\rm eff}(t) = \frac{\log S(0,t)-\log S(0,t-1)}{\log t-\log(t-1)} \ .
\end{equation}
A more detailed discussion of each dimension is in order at this point:

$d = 1$. The transverse spatial directions become obsolete, and with the exact
renormalization group result $\Delta = 1/3$ we obtain $z_\parallel = 3/2$ and 
$\zeta = 1 / z_\parallel = 2/3$.
As mentioned above, the scaling behavior for the equivalent surface growth 
model falls into the KPZ universality class \cite{kpz}.
The central quantity in this description is the height-height correlation 
function
\begin{equation}
  \label{hhcorr}
  C(x,t) = \Big\langle [h(x,t)-h(0,0)]^2 \Big\rangle ,
\end{equation}
which is related to the density-density correlation function via
\cite{spohn_proc}
\begin{equation}
  \label{hhddrelation}
  S(x,t) = \frac{1}{8} \frac{\partial^2}{\partial x^2} C(x,t) .
\end{equation}
Asymptotically, it takes the scaling form
\begin{equation}
\label{scalehh}
  C(x,t) = b^{- 2 \alpha} C(bx, b^{z_\parallel} t) , 
\end{equation}
with the roughness exponent $\alpha$ (note that in the growth model literature,
the dynamical exponent $z_\parallel$ is usually labeled $z$).
According to Eqs.~(\ref{hhddrelation}) and (\ref{scalehh}), we can therefore 
identify $2 - 2 \alpha = (d + \Delta) z_\parallel / z = 1$ in one dimension, or
$\alpha = 1/2 = 2 - z_\parallel$ \cite{Forster,kpz,Jan}.

$d = d_c = 2$. At the upper critical dimension, logarithmic corrections arise.
The simple power law (\ref{zeta}) is replaced with \cite{Forster,Jan}
\begin{equation}
  S(0,0,t) \sim \frac{1}{t (\log t)^{1/2}} \ .
\end{equation}
Accordingly, in order to capture the approach to the asymptotic scaling limit 
we need to redefine the local exponent in this case as
\begin{equation}
  \label{2dloc}
  S(0,0,t) (\log t)^{1/2} \propto t^{-\zeta_{\rm eff}(t)} .
\end{equation}
It turns out though that the mean-field approximation 
\begin{equation}
  \label{MF}
  S_{\rm MF}(0,0,t) \sim t^{-1} 
\end{equation}
provides reasonable agreement with our simuation data as well.

$d > 2$. Above the critical dimension, the mean-field predictions for the
scaling exponents $\Delta = 0$ and $z = 2 = z_\parallel$, $\zeta = d/2$ should 
provide an adequate description of the density auto-correlation function data.

Finally, we address the two-time density auto-correlation function $S(0,t,s)$ 
in the non-stationary regime when the system relaxes from a strongly correlated
initial state that considerably differs from any non-equilibrium steady-state
configuration.
As we shall explain in more detail in Sec.~IV.B below, one expects that the 
physical aging scaling regime $s \ll t$ is governed by the simple aging scaling
form \cite{pleimbook}
\begin{equation}
  \label{agescaledds}
  S(0,t,s) = s^{-\zeta} h_S(t/s) .
\end{equation}
Indeed, we shall see that this scaling ansatz yields satisfactory data 
collapse.
We remark that in the literature on glassy systems, an alternate so-called
subaging fitting, $S(0,t,s) = h_S(t/s^\mu$) with $\mu < 1$, has been popular 
\cite{pleimbook}.

\section{Steady-state results}
\label{steady}

\subsection{Recurrence oscillations in small systems}

\begin{figure}[b] 
  \includegraphics[width=\columnwidth,clip]{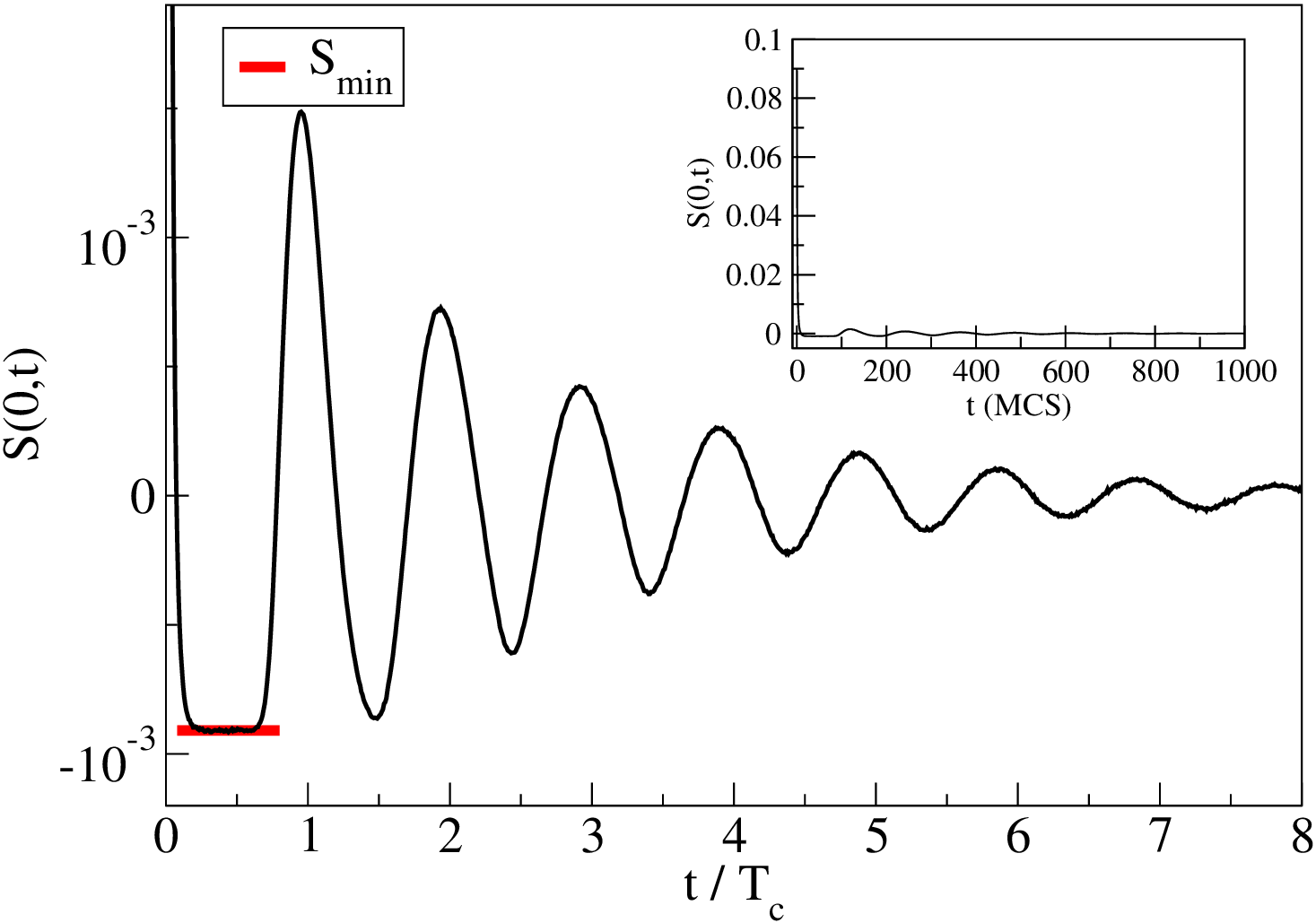}
  \caption{({\it Color online.}) Characteristic recurrence oscillations in the
  density auto-correlation function $S(0,t)$ for a one-dimensional driven 
  lattice gas (TASEP) at low density $\rho = 0.1$ and $L = 100$ (close-up).
  The horizontal (red) line indicates the minimum value 
  $S_{\rm min} = - 0.909 \times 10^{-3}$ (red line), see Eq.~(\ref{smin}).  
  The inset has been included to show the actual magnitude of the oscillations.
  The return time is $T_c = 125$ MCS.  
  The data are averaged over 8,000,000 realizations. \label{osc1}}
\end{figure}
\begin{figure}[t]
  \includegraphics[width=\columnwidth,clip]{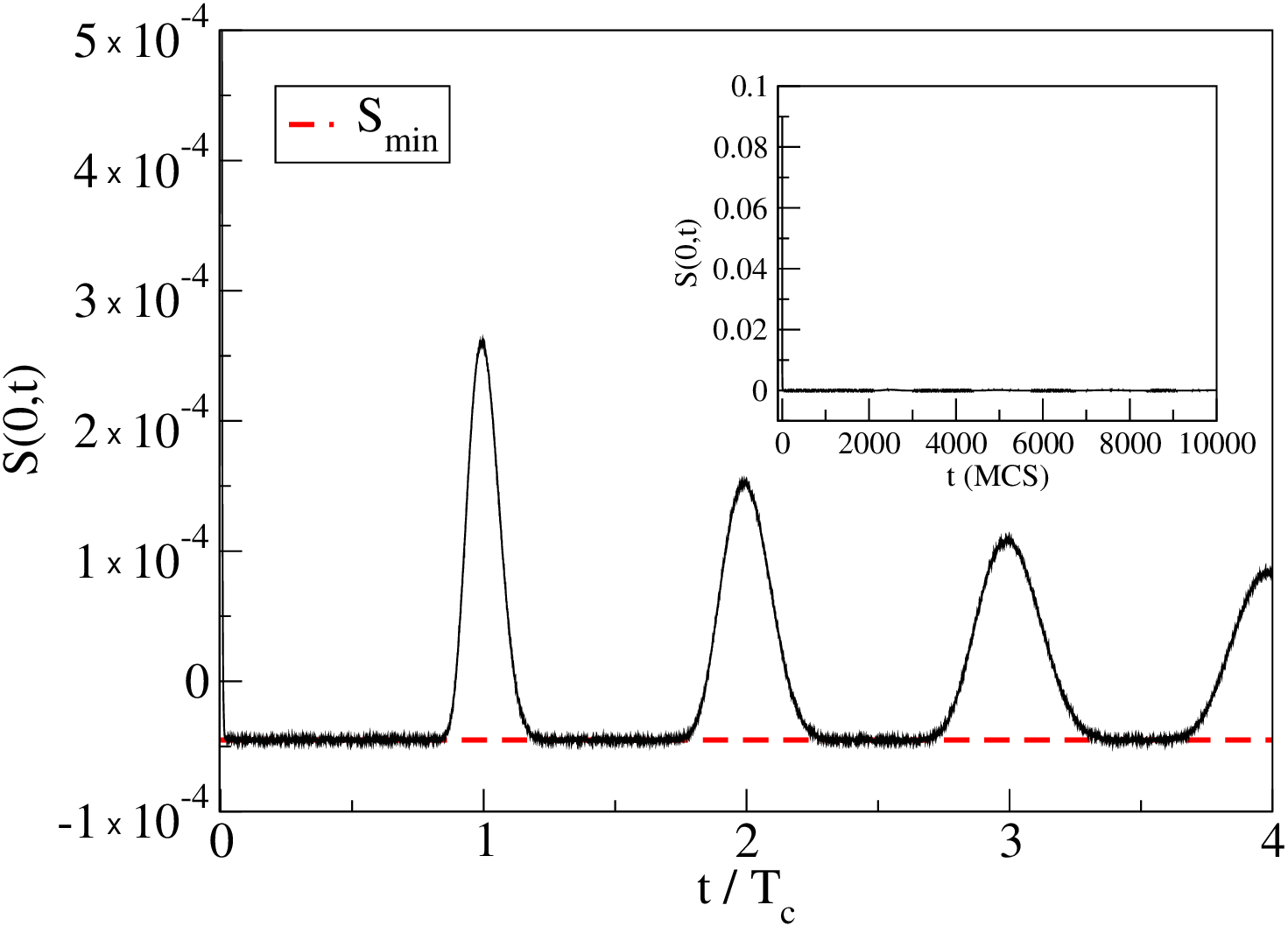}
  \caption{({\it Color online.}) Recurrence oscillations in the driven lattice 
  gas (TASEP) auto-correlation function $S(0,t)$, again for $\rho = 0.1$, but a
  larger one-dimensional ring of length $L = 2000$; here 
  $S_{\rm min} = 0.45 \times 10^{-4}$, indicated by the dashed (red) line. 
  The return time is $T_c = 2500$ MCS, and the data are averaged over 
  8,000,000 realizations. 
  \label{osc2}}
\end{figure}

We begin with a discussion of small-scale properties of the driven lattice gas 
on a one-dimensional ring of length $L$ (TASEP).  
When the total density $\rho = N/L \not= 1/2$, collective density fluctuations
travel around the system with mean velocity $v_c = 1 - 2 \rho$.
This causes characteristic oscillations in the time-dependent density 
auto-correlation function $S(0,t)$, with peaks at multiples of the return time
$T_c = L / v_c$, as can be seen clearly in Figs.~\ref{osc1} and \ref{osc2}, 
which show data obtained for $L = 100$ and $L = 2000$, respectively, both at 
low density $\rho = 0.1$.
(We only report results for densities less than $1/2$, since as a consequence 
of particle-hole symmetry, the results for $\rho = 0.5 - \delta$ are equivalent
to those for $\rho = 0.5 + \delta$.)

Observe the flat section in Fig.~\ref{osc1} at a negative value 
$S_{\rm min} = - 0.9 \times 10^{-3}$ in the first minimum.
Invoking the fact that in the steady state, all configurations carry the same 
statistical weight, we can understand this lower bound on the density
auto-correlation function as follows:
Consider a single occupied site $x$ at time $t$, hence $n(x,t) = 1$.
At time $t'>t$, assume that the particle has left the site $x$, but has not 
yet traversed the entire system and returned: this situation will give the
lowest average possible value for the correlation function.
While the new location of the particle previously at $x$ can be inferred from
the tagged particle velocity $v_t$, Eq.~(\ref{vtag}), the other $N-1$ particles
are left to fill any of the remaining $L-1$ sites with equal likelihood, which 
gives a probability $(N-1)/(L-1)$ for site $x$ to become occupied at $t'$.
Averaging over all $L$ sites, we thus arrive at the same average minimum value
(\ref{smin}) for the auto-correlation function $S(0,t > 0)$ when the original 
particle has traveled away that we derived earlier for the equal-time density 
cumulant for $S(\vec x' - \vec x \not = 0, t = 0)$:
After a brief decorrelation time, typical configurations at any given site at 
later time $t'$ are statistically correlated to the reference configuration at
$t<t'$ in the same manner as different sites are at equal time in the 
stationary state.
The amplitude of the recurrence oscillations naturally decreases with 
increasing system size.
Note, however, that the mean negative value $S_{\rm min} = 0.45 \times 10^{-4}$
is still clearly visible in Fig.~\ref{osc2} for $L = 2000$.

\begin{figure}[t]
  \includegraphics[width=\columnwidth,clip]{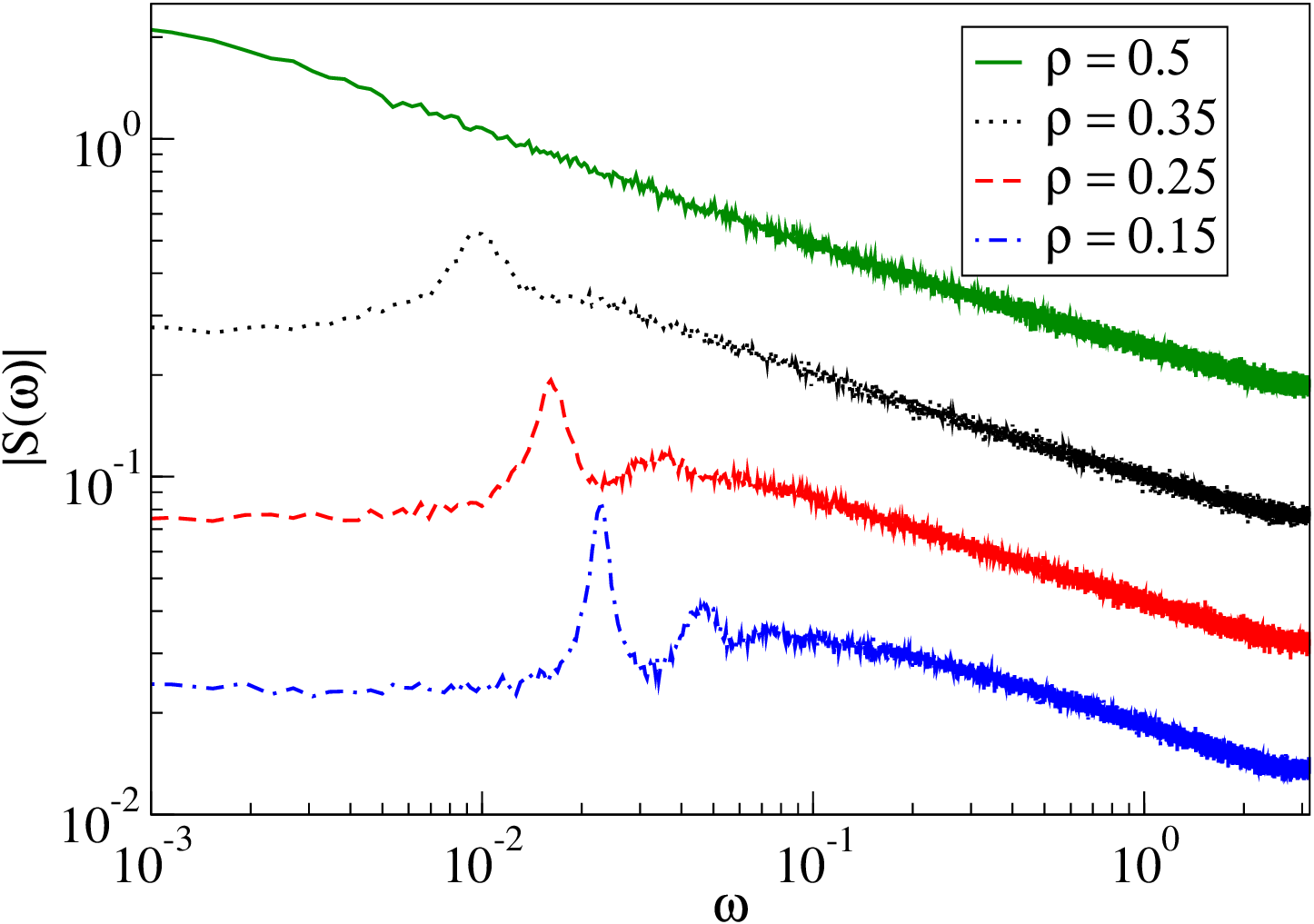}
  \caption{({\it Color online.}) Fourier transform $|S(\omega)|$ of $S(0,t)$ 
  for one-dimensional driven lattice gases (TASEP) with $L = 400$, at various 
  densities $\rho = 0.15$, $0.25$, $0.35$, and $0.5$.
  The curves are vertically shifted to separate them and show the peaks.
  In each case, the data were averaged over 800,000 realizations. \label{fft1}}
\end{figure}
In order to extract the period of the recurrent density fluctuations, we 
Fourier transform the auto-correlation function, 
$S(\omega) = \int S(0,t) e^{i \omega t} dt$.
The first peak in $|S(\omega)|$ is caused by the movement of a density 
fluctuation at a frequency $\omega_c = 2 \pi / T_c$, see Fig.~\ref{fft1}.
At low densities, higher harmonics are also visible.
The characteristic frequency $\omega_c$ vanishes as the density approaches 
$1/2$.
We finally mention previous work by Adams {\em et al.} \cite{adams} and Cook 
and Zia \cite{cook} who studied density oscillations in the open TASEP that are
also caused by fluctuations that traverse the entire system.  
Gupta {\em et al.} studied similar oscillations in the displacement variance 
for tagged particles \cite{oscillate}.  
Exact calculations of the spectral gap yield evidence of oscillations as well 
\cite{spec_osc2,spec_osc1,kirone_rev}, which are also found in exact solutions
for the time-dependent conditional probabilities and single particle 
correlation function \cite{exact3}.

\subsection{Finite-size scaling}

\begin{figure}[t]
  \includegraphics[width=\columnwidth,clip]{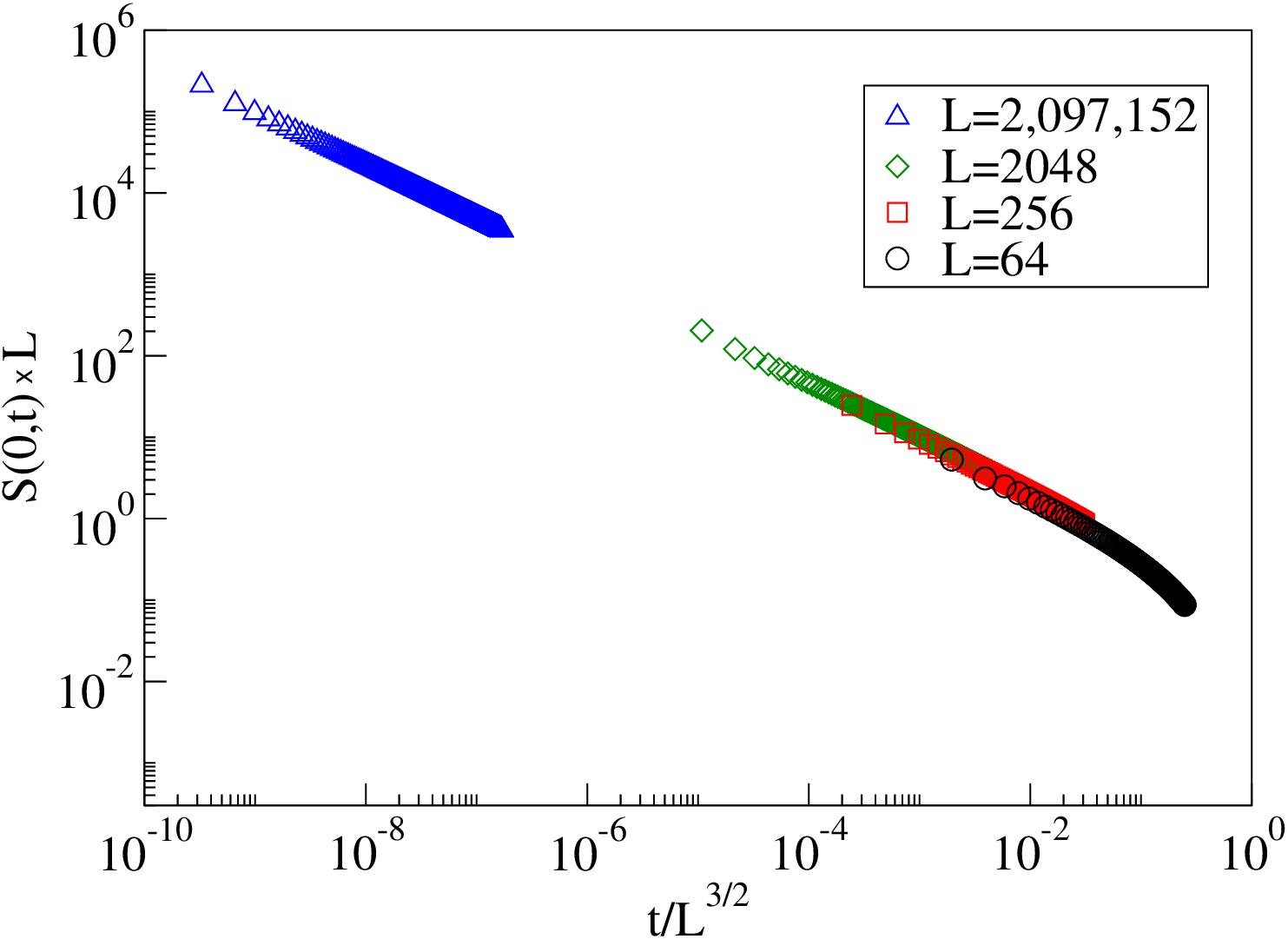}
  \caption{({\it Color online.}) Finite-size scaling for the density
  auto-correlation function for $L = 64$ (data averaged over 80,000 
  realizations), $L = 256$ (10000 realizations), $L = 2048$ 
  (40,000 realizations), and $L = 2,097,152$ (60 realizations). 
  \label{fss1}}
\end{figure}
Next we analyze the finite-size scaling properties of the density 
auto-correlation function for the driven lattice gas with exclusion in one 
dimension (TASEP).
The expected finite-size scaling form is readily obtained from the general 
expression (\ref{xtgen}) by setting $L = b^{-(1+\Delta)}$, and omitting the
transverse space directions:
\begin{equation}
  S(x,t) = L^{-1} S_{\rm FS}\Bigl( x/L, t / L^{z_\parallel} \Bigr) .
\end{equation}
Taking $x \to 0$ we find for the auto-correlation function
\begin{equation}
  \label{fssc}
  S(0,t) = L^{-1} S_{\rm FS}\Bigl( t / L^{z_\parallel} \Bigr) .
\end{equation}
As shown in Fig.~\ref{fss1}, the finite-size scaling form (\ref{fssc}) produces
satisfactory data collapse with $z_\parallel = 3/2$ for our simulation results 
for vastly different lattice sizes ranging from $L = 64$ to $L = 2,097,152$.
At $t / L^{z_\parallel} \approx 0.1$ one observes the expected finite-$L$
cutoff from the scaling form (\ref{fssc}), as the return time $t_r$ is 
approached.
We note that earlier work has studied the finite-size scaling properties of the
TASEP with open boundaries in the maximum current phase which is equivalent to 
the periodic case; however much smaller system sizes of respectively $L = 50$ 
and $L = 257$ were utilized by Pierobon {\em et al.} \cite{Pier} and Juh\'asz 
and Santen \cite{Juh}.

\subsection{Long-time properties}

\subsubsection{Scaling functions}

\begin{figure}[t]
  \includegraphics[width=\columnwidth,clip]{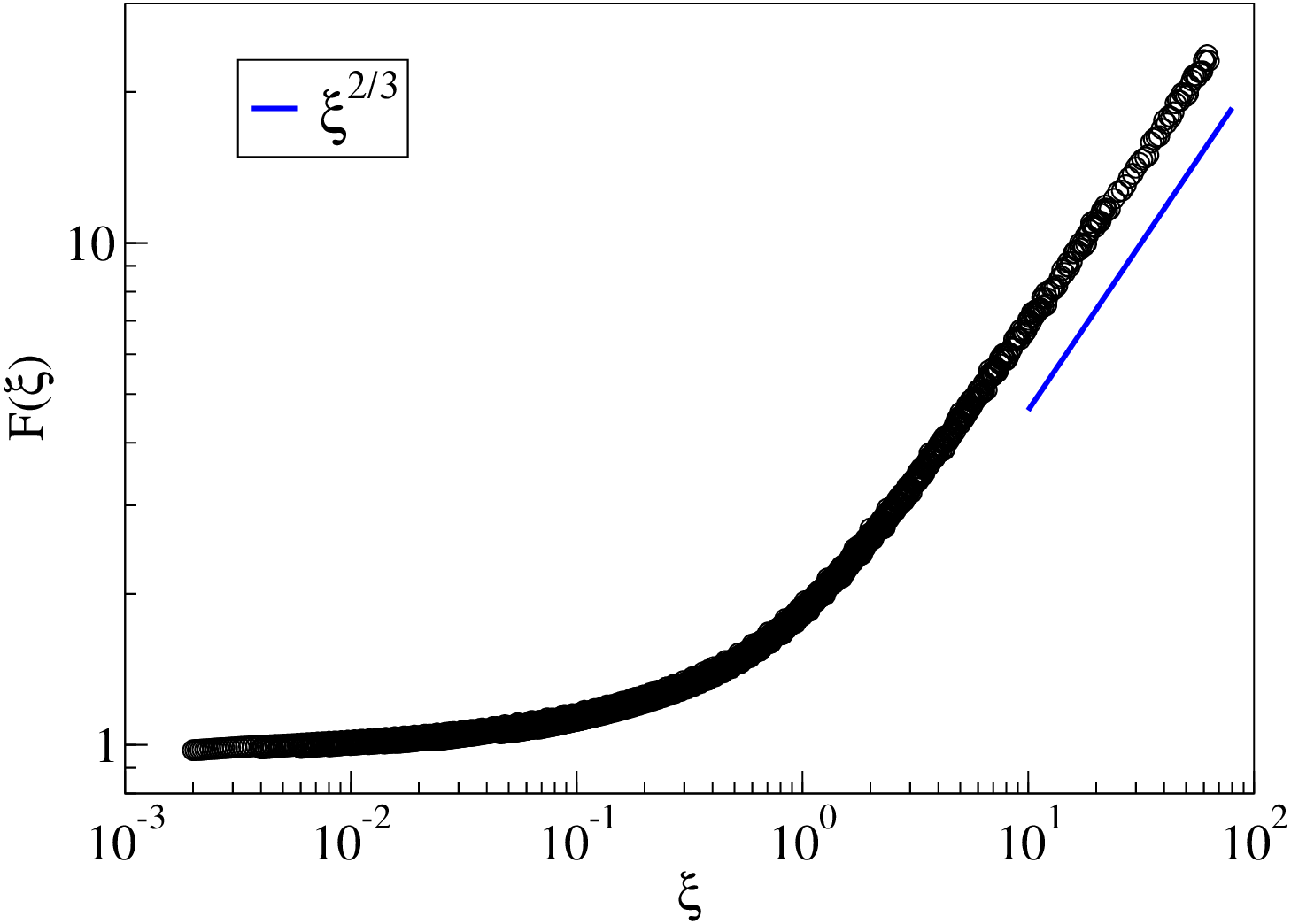}
  \caption{({\it Color online.}) Double-logarithmic plot of the scaling 
  function $F(\xi)$, defined in Eq.~(\ref{esc}), obtained from Monte Carlo 
  simulations in a driven lattice gas on a ring of length $L = 2000$, with 
  $x \in [0,64]$ and $t \in [0,64]$.
  The data are averaged over 200 realizations. \label{uwescale}}
\end{figure}
\begin{figure}[b] 
  \includegraphics[width=\columnwidth,clip]{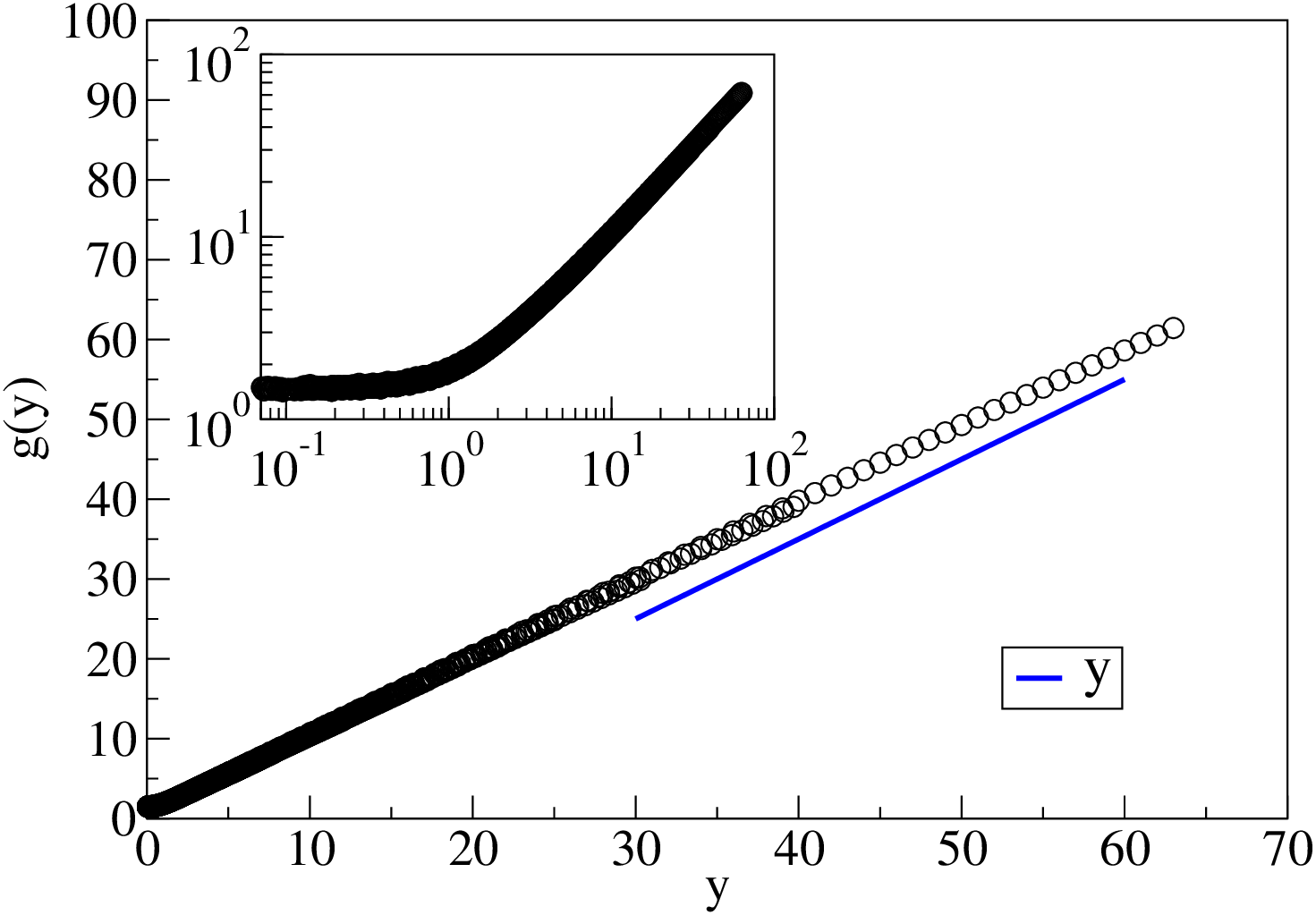}
  \caption{({\it Color online.}) The scaling function $g(y)$ defined in 
  Eq.~(\ref{ssc}), from the same data as in Fig.~\ref{uwescale} ($L = 2000$, 
  $x \in [0,64]$, and $t \in [0,64]$, averaged over 200 realizations).
  The inset represents the same data in a double-logarithmic plot to display 
  the crossover to a constant as $y \to 0$.
  \label{spohnscale}}
\end{figure}
We now address the stationary height-height correlation function (\ref{hhcorr})
in one dimension, and explicitly determine associated scaling functions.
Setting $b = 1 / |x|$, and recalling $\alpha = 1/2$ and $z_\parallel = 3/2$,
Eq.~(\ref{scalehh}) becomes \cite{uwehwa}
\begin{equation}
  \label{esc}
  C(x,t) \propto |x| F(\xi) , \ \xi = t / |x|^{3/2} \ ,
\end{equation}
whereas equivalently the matching condition $b = t^{-1/z_\parallel}$ leads to 
\cite{spohnlong}
\begin{equation}
  \label{ssc}
  C(x,t) \propto t^{2/3} g(y) , \ y = x / t^{2/3} \ .
\end{equation}
Consistency then requires that 
\begin{equation}
   \label{larg}
   F(\xi) \propto \xi^{2/3} , \ g(y) \propto |y| 
\end{equation}
for large arguments $\xi \gg 1$ and $y \gg 1$.
Figures~\ref{uwescale} and \ref{spohnscale} depict the scaling functions 
$F(\xi)$ and $g(y)$ as obtained from our simulations for a driven periodic 
lattice gas with exclusion on a ring of length $L = 2000$.
The Monte Carlo data nicely confirm the relations (\ref{larg}).
The crossover from $F(\xi) = \rm const.$ for small arguments $\xi$ to the 
asymptotic behavior $\sim \xi^{2/3}$ visible in Fig.~\ref{uwescale} can be 
compared with the numerical solution of the mode-coupling equations depicted in
Fig.~2 of Ref.~\cite{uwehwa} and the simulation results shown in Fig.~2 of
Ref.~\cite{monte_1}.
The mode-coupling approximation appears to predict sharper crossover features,
with $F(\xi)$ remaining constant until $\xi \approx 0.5$, where our data 
already indicate a noticeable curvature in the scaling function $F$.  
The simulation data in Ref.~\cite{monte_1}, obtained for much larger systems
with $L = 1048576$, also display sharper crossover features than our data for
$L = 2000$.

\begin{figure}[t] 
  \includegraphics[width=\columnwidth,clip]{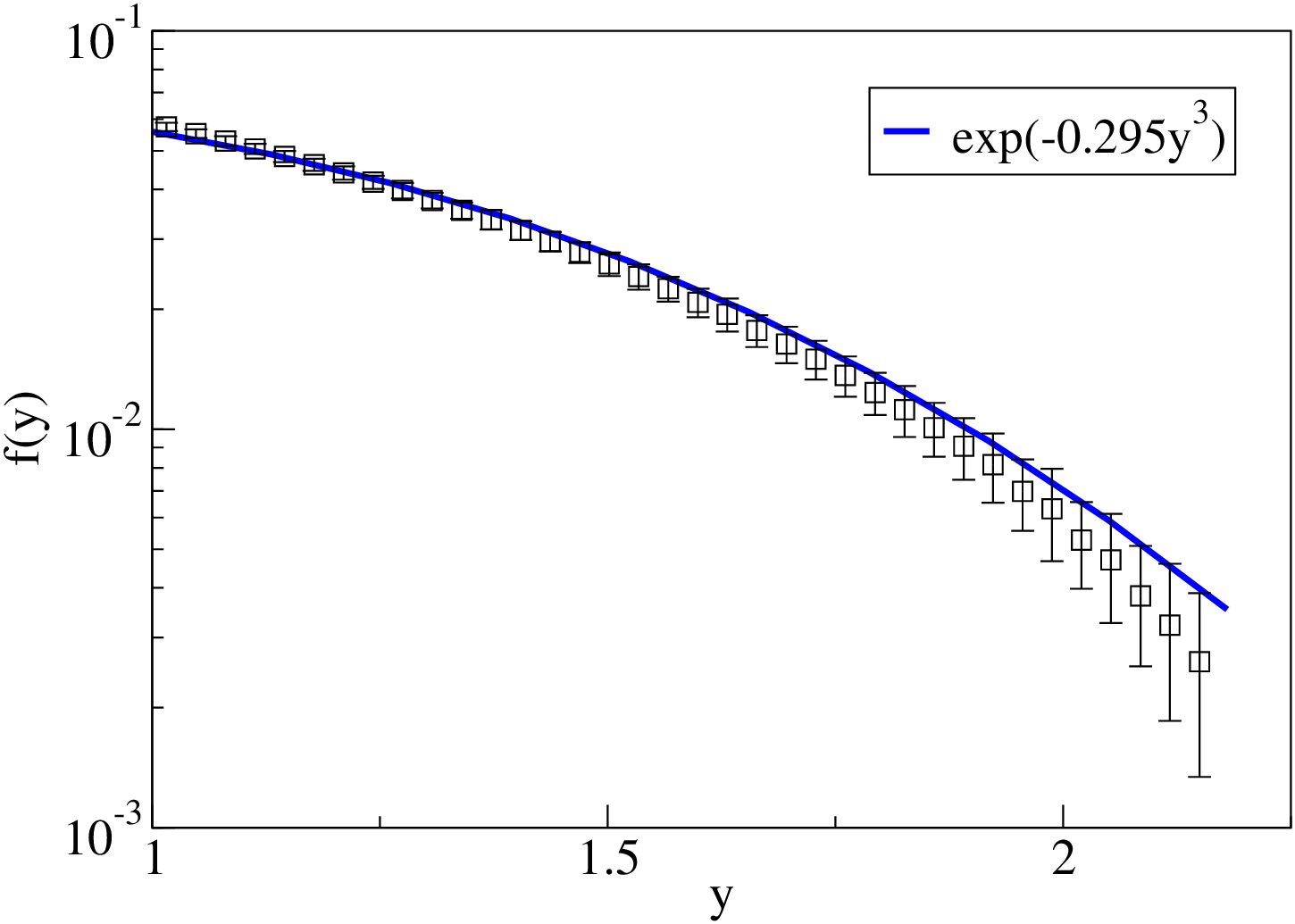}
  \caption{({\it Color online.}) Logarithmic plot of the scaling function 
  $f(y)$, defined in Eq.~(\ref{ssca}), for $L = 2000$, with $x \in [0,256]$ and
  $t \in [0,256]$, with the data averaged over 400,000 realizations. 
  The error bars indicate one standard deviation. \label{gpp}}
\end{figure}
Combining Eqs.~(\ref{hhddrelation}) and (\ref{ssc}), we recover the scaling
form (\ref{scaledt}) for the density-density correlation function in one 
dimension,
\begin{equation}
  \label{ssca}
  S(x,t) \propto t^{-2/3} f(y) . 
\end{equation}
Our Monte Carlo simulation data for the scaling function $f(y)$, measured by
averaging over a very large number of realizations, are plotted in 
Fig.~\ref{gpp}.
Within the error bars, our data are in reasonable aggreement with the stretched
exponential function $f(y) \propto \exp(-c y^3)$, where $c = 0.295(5)$, that 
was computed in Ref.~\cite{spohnlong}.

\subsubsection{Local exponent}

\begin{figure}[t]
  \includegraphics[width=\columnwidth,clip]{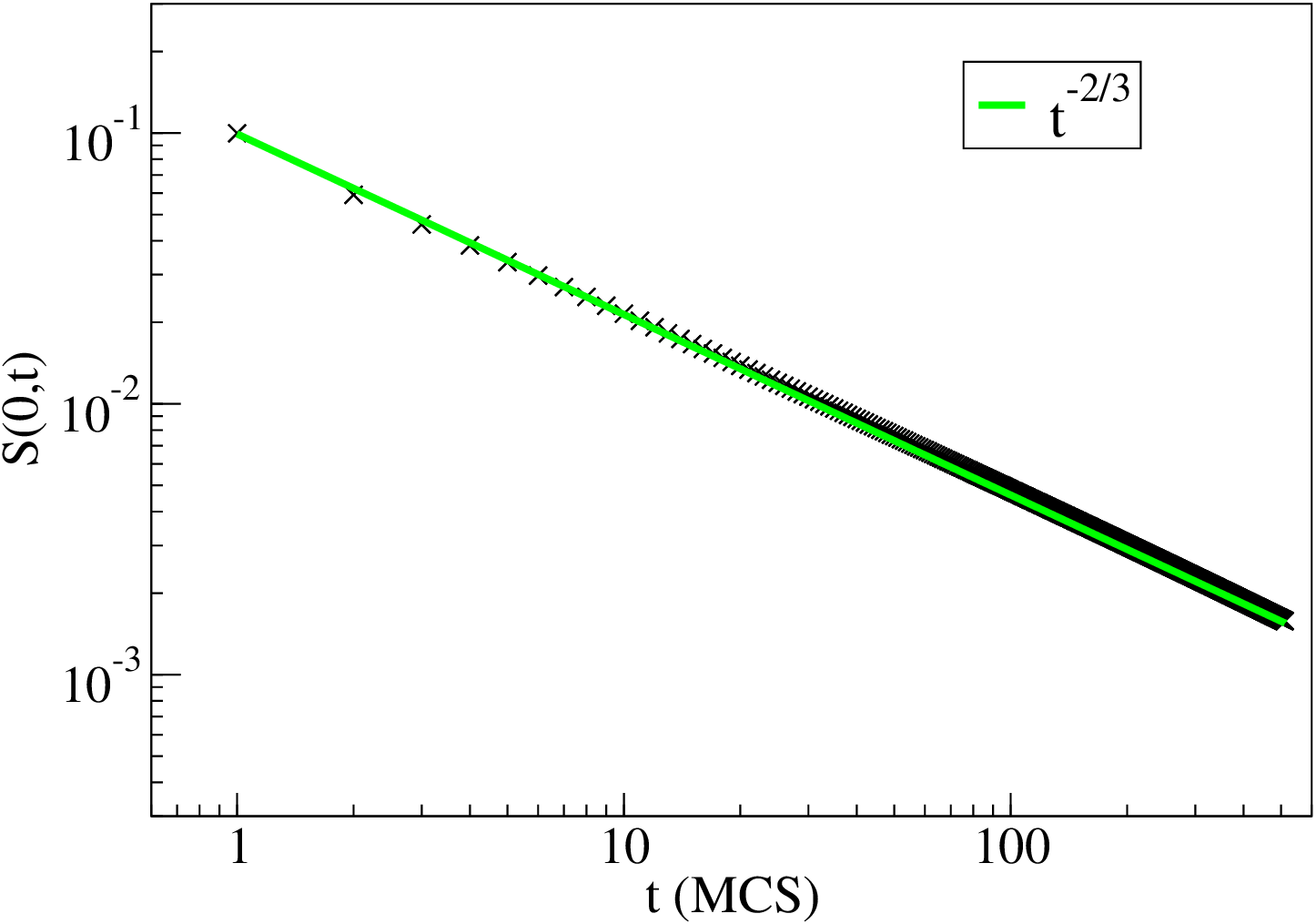}
  \caption{({\it Color online.}) Double-logarithmic plot of the stationary 
  auto-correlation function $S(0,t)$ for a one-dimensional driven lattice gas 
  of length $L = 2048$ with exclusion.
  The data are averaged over 40,000 realizations. 
  The light green line represents a power law with exponent $- 2/3$.  
  \label{fig:lnln}}
\end{figure}
In Fig.~\ref{fig:lnln} we plot the density auto-correlation function $S(0,t)$ 
as obtained from our simulations for a one-dimensional driven lattice gas with 
exclusion with $L = 2048$ (see also Fig.~\ref{fss1}).
From Eq.~(\ref{zeta}) with $\zeta = 1  / z_\parallel = 2/3$, we expect the 
asymptotic long-time power law $S(0,t) \sim t^{-2/3}$, which fits the Monte
Carlo data well for $t > 100$ on the double-logarithmic plot.
The approach to the expected asymptotic power law auto-correlation decay can be
carefully probed by studying the local effective exponent $\zeta_{\rm eff}(t)$ 
defined in Eq.~(\ref{eq:loc}).
\begin{figure}[b]
  \includegraphics[width=\columnwidth,clip]{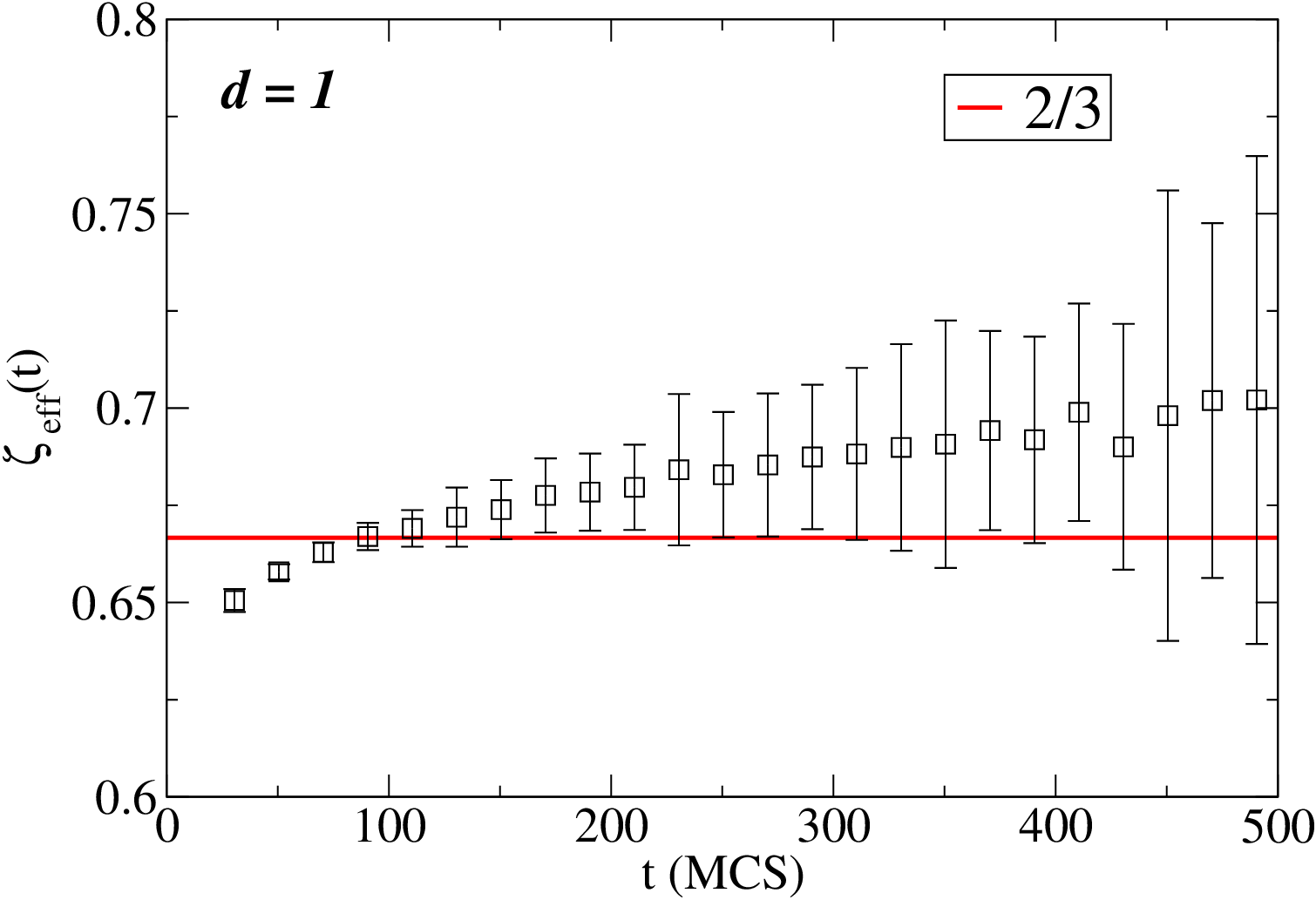}
  \caption{({\it Color online.}) The local auto-correlation exponent 
  (\ref{eq:loc}) obtained from Monte Carlo simulation data for a 
  one-dimensional driven lattice gases with length $L = 2048$ and for 
  $\rho = 1/2$.  
  The data are averaged over 40,000 realizations.  
  The error bars represent one standard deviation, and the (red) horizontal 
  line the asympotic value $2/3$. 
\label{fig:loc2048}}
\end{figure}

\begin{figure}[t]
  \includegraphics[width=\columnwidth,clip]{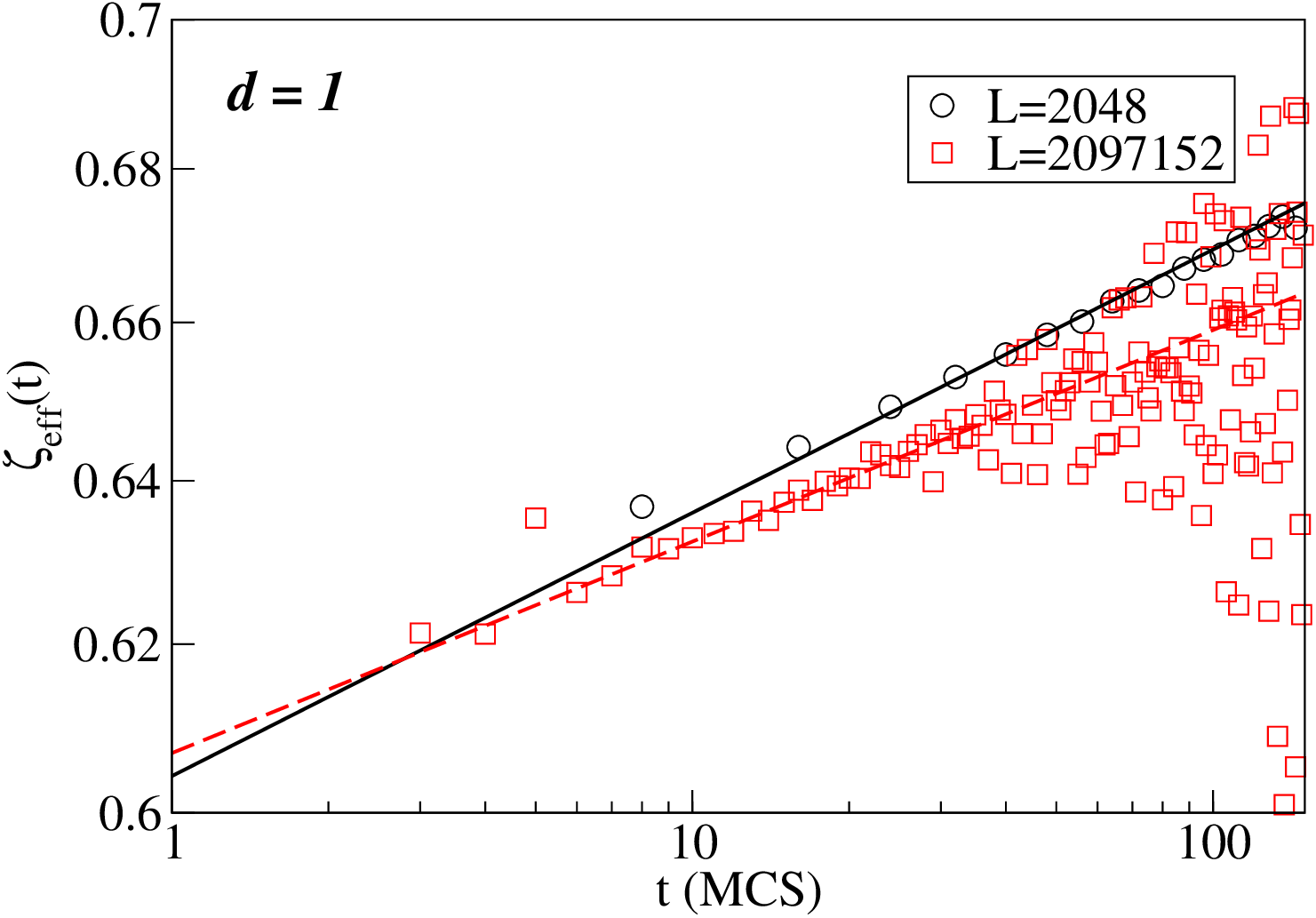}
  \caption{({\it Color online.}) Time dependence of the local exponent 
  $\zeta_{\rm eff}(t)$ for one-dimensional driven lattice gases with $L = 2048$
  (black circles, data averaged over 40,000 realizations) and $L = 2097152$ 
  (red squares, 60 realizations). \label{2sizelog}}
\end{figure}
\begin{figure}[b]
  \includegraphics[width=\columnwidth,clip]{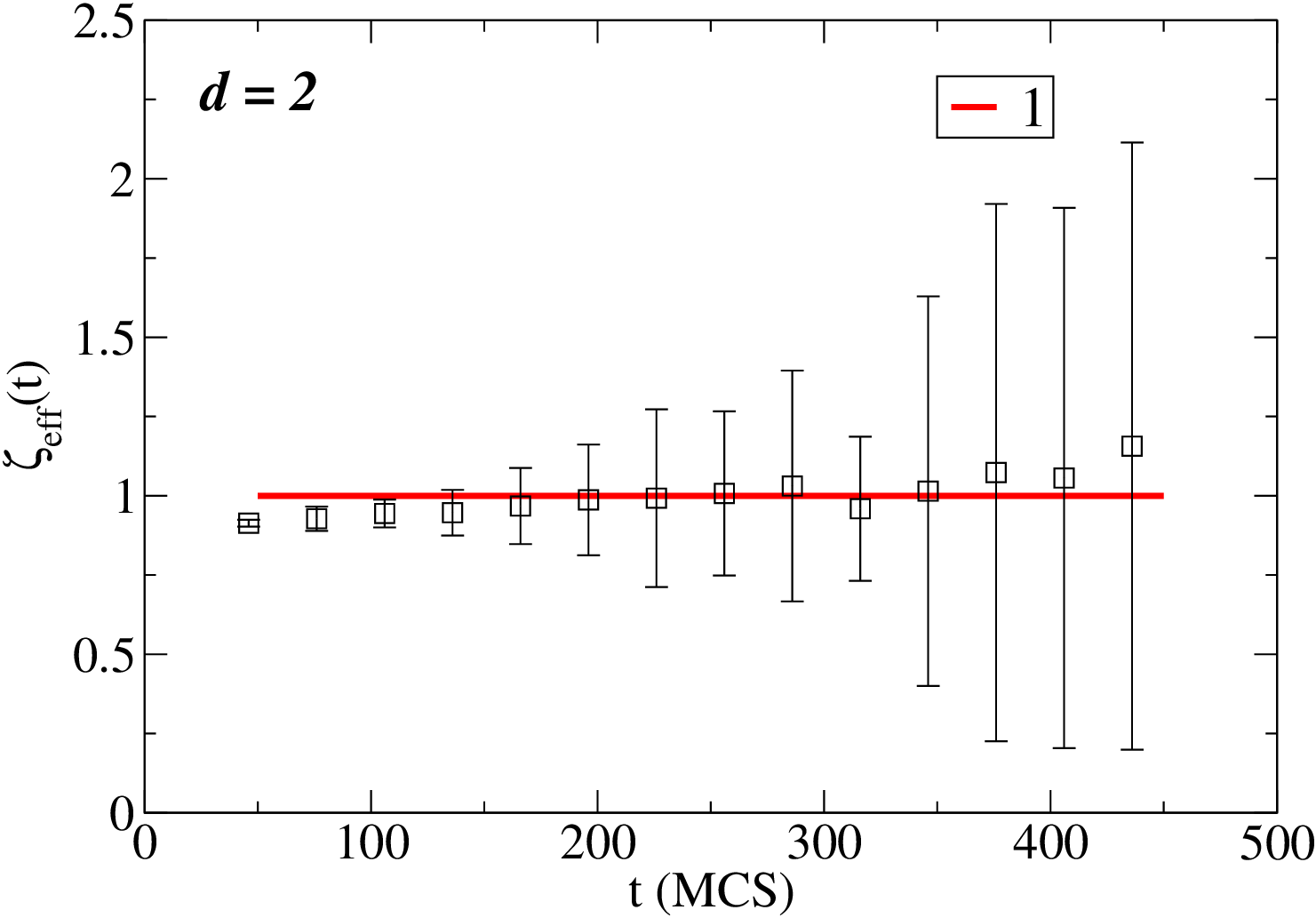}
  \caption{({\it Color online.}) The effective exponent $\zeta_{\rm eff}(t)$, 
  Eq.~(\ref{2dloc}), for a two-dimensional driven lattice gas with 
  $L_x = L_y = 100$ and $\rho = 1/2$.  
  The data are averaged over 600,000 realizations.
  The error bars indicate one standard deviation, and the (red) horizontal line
  the asympotic value $1$. \label{2dexp}}
\end{figure}
Figure~\ref{fig:loc2048} shows that the local exponent increases with time and
in fact surpasses the asymptotic value $2/3$, indicated by the horizontal line.
In order to demonstrate that this is actually a finite-size effect, albeit a 
quite unusual one, we display in Fig.~\ref{2sizelog} a comparison of simulation
data for the time dependence of the effective exponent for systems of sizes 
$L = 2048$ and $L = 1024 \times 2048 = 2097152$ on a logarithmic time scale.
Even in the much larger lattice, $\zeta_{\rm eff}(t)$ still rises beyond the 
asymptotic exponent value $\zeta = 2/3$ of the infinite system, yet at later 
time.  
Since we have not observed this overshooting in unbiased lattice gases with
exclusion, it is clearly a consequence of the non-equilibrium drive.  
We note that the extremely slow crossover towards the asymptotic temporal power
law has been recorded before in a model of reconstituting dimers that can be 
mapped to the TASEP \cite{MBarma}. 

\begin{figure}[t]
  \includegraphics[width=\columnwidth,clip]{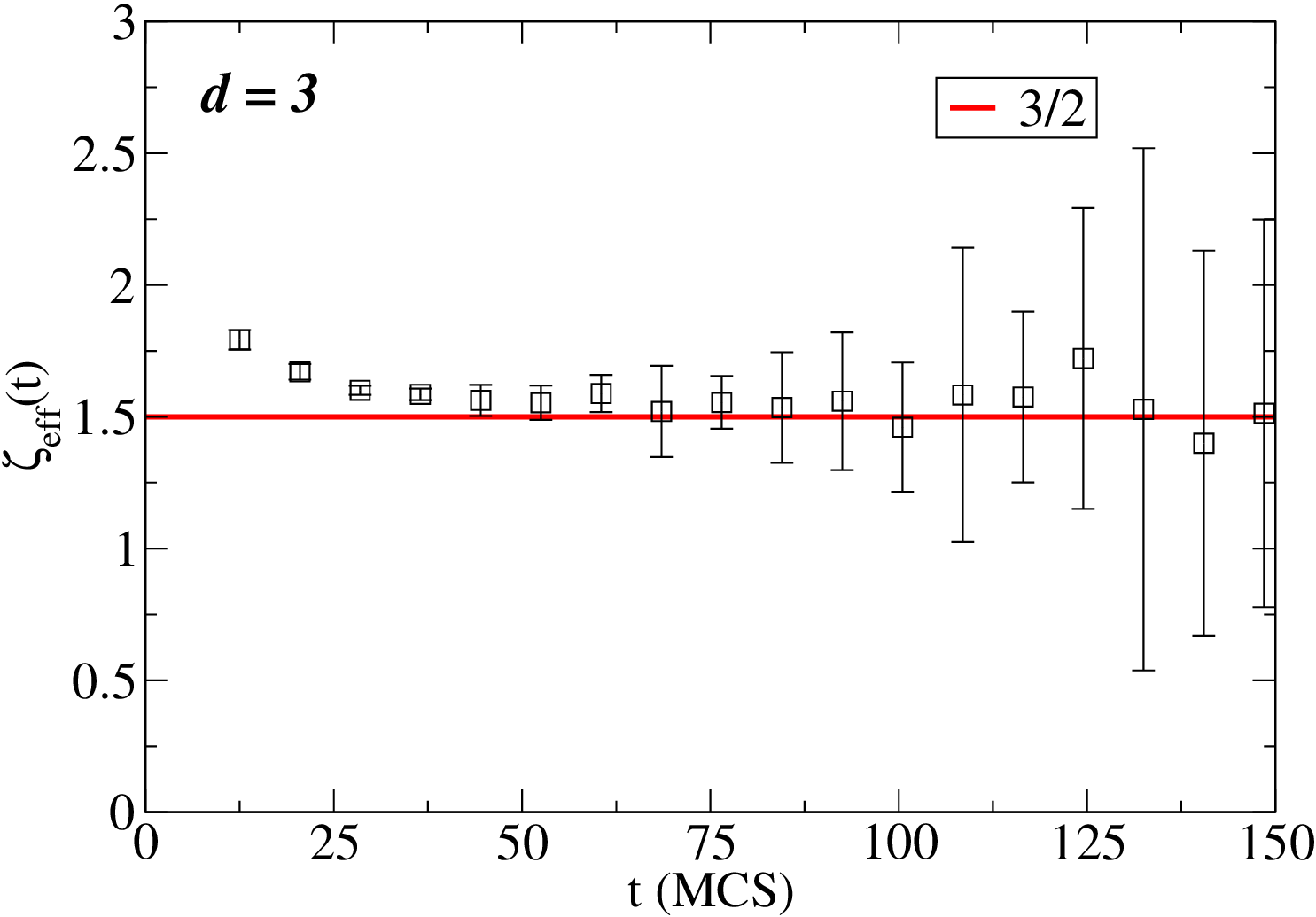}
  \caption{({\it Color online.}) The local exponent $\zeta_{\rm eff}(t)$ for a
  three-dimensional driven lattice gas with $L_x = L_y = L_z = 50$ and 
  $\rho=1/2$.  
  The data are averaged over 1,320,000 realizations.  
  The error bars indicate one standard deviation, and the (red) horizontal line
  the asympotic value $3/2$. \label{3dexp}}
\end{figure}

We next compare the behavior of the local auto-correlation exponent 
$\zeta_{\rm eff}(t)$ for the one-dimensional driven lattice gas with its
two- and three-dimensional counterparts, see Figs.~\ref{2dexp} and \ref{3dexp}.
In each case, the system is half-filled ($\rho = 1/2$), and the hopping rates 
are equal in each direction transverse to the drive, but set totally asymmetric
parallel to the drive.  
For a two-dimensional driven lattice gas ($L_x = L_y = 100$), we depict the
effective exponent as defined in Eq.~(\ref{2dloc}) in Fig.~\ref{2dexp}, and
analogous results for the three-dimensional case are shown in Fig.~\ref{3dexp}.
Since we are at or above the critical dimension $d_c = 2$, respectively, the
asymptotic value is $\zeta = d/2$.
Indeed, the local exponent $\zeta_{\rm eff}(t)$ quickly relaxes to $1$ for 
$d = 2$ and $3/2$ for $d = 3$, and never surpasses these expected asymptotic 
values.
The anomalously slow convergence and unusual finite-size behavior in one
dimension are consequently caused by the strong out-of-equilibrium correlations
present only for $d < d_c$.

\begin{figure}[t] 
  \includegraphics[width=\columnwidth,clip]{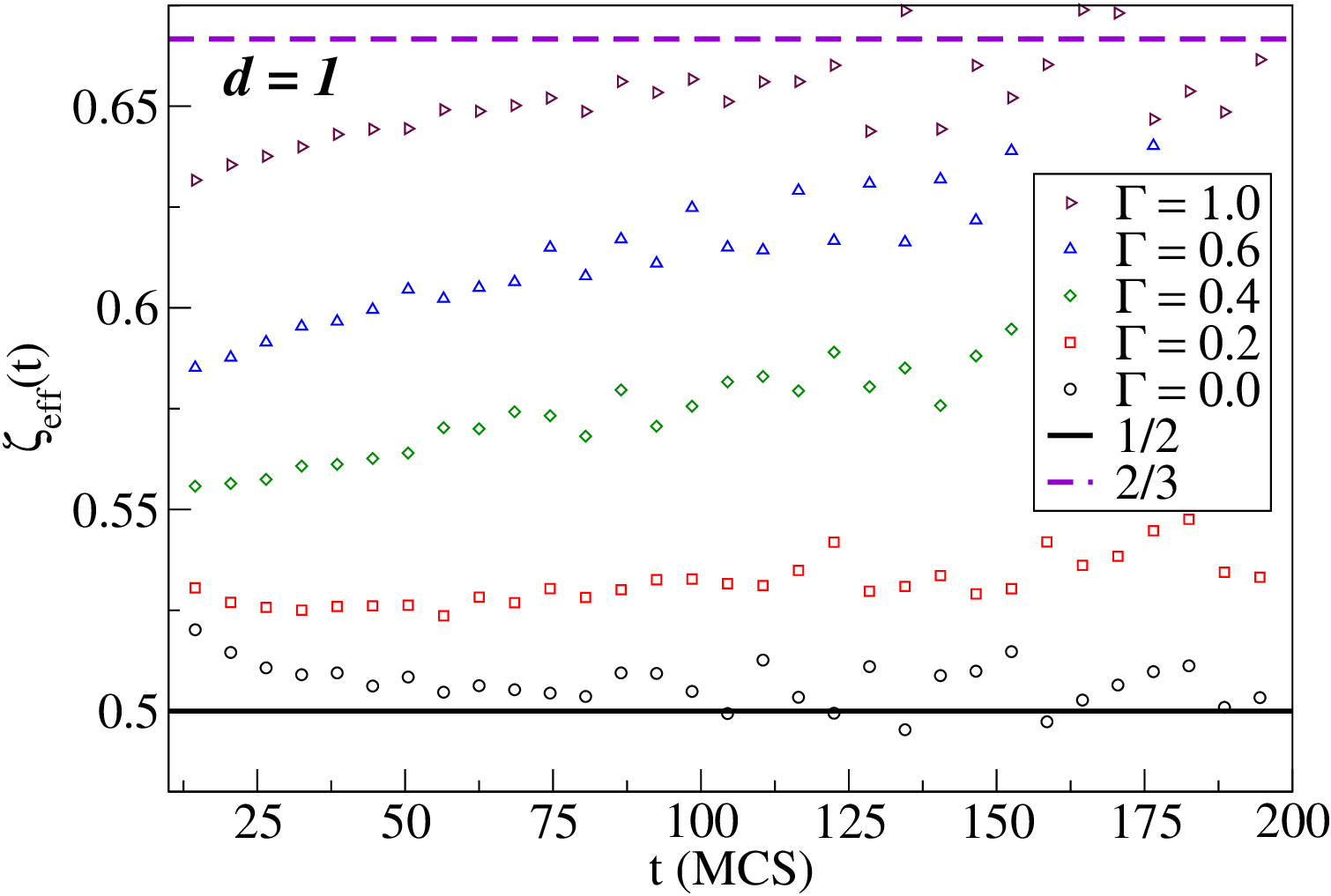}
  \caption{({\it Color online.}) Local auto-correlation exponent 
  $\zeta_{\rm eff}(t)$ for one-dimensional driven lattice gases (ASEP) with 
  different hopping biases $\Gamma = 1.0$ (TASEP), $0.6$, $0.4$, $0.2$ (all 
  ASEP), and $0.0$ (vanishing drive, SEP); $L = 4096$ and $\rho = 1/2$. 
  The data are averaged over 5,000,000 realizations each. \label{fig:asep}}
\end{figure}

Up to this point, we have provided results only for driven lattice gases with
zero backward hopping rate along the drive direction (TASEP in one dimension).
Assigning the forward hopping probability $p$ and backward probability $q$, we 
can define the bias $\Gamma = p - q = 2 p - 1$.
As long as $1/2 < p < 1$ and $0 < \Gamma < 1$ (ASEP), a macroscopic particle
current persists, the system remains out of equilibrium, and one expects 
qualitatively the same behavior as for maximum bias or drive, $p = \Gamma = 1$.
Thus, at long times, the temporal decay of the density auto-correlation 
function should be described by a power law with exponent $\zeta = 2/3$ in one
dimension, for any non-zero value of the hopping bias $\Gamma$.  
However, note that this statement is valid at asympotically long times for
sufficiently large systems, and a simple power-law decay will not necessarily 
be observable in finite systems with weak bias.
Monte Carlo simulation data for one-dimensional driven lattice gases with
various bias values are shown in Fig.~\ref{fig:asep}.
As can be seen, the local exponent $\zeta_{\rm eff}(t)$ relaxes increasingly
slowly towards its asymptotic value $2/3$ as the bias is reduced.
The figure illustrates that for low bias and at short times, the density 
auto-correlation decay rather follow the symmetric exclusion process (SEP)
relaxation, for which $\zeta = 1/2$.  
For example, for $\Gamma = 0.2$ a distinct decrease in the local exponent is 
observed for $t < 30$, later followed by a very slow increase.  

\section{Non-equilibrium relaxation}
\label{aging}

\subsection{Transitions between steady states}

Let us finally consider transient relaxation properties of driven lattice gases
with exclusion.
For example, starting from a steady state obtained with a specified set of 
hopping probabilites $p$ and $q = 1 - p$, we may wish to understand how the 
system responds to a sudden change of these probabilities to new values $p'$ 
and $q' = 1 - p'$.  
However, recall that the non-equilibrium steady state probability distribution
(\ref{probC}) is independent of the hopping probabilities.  
This means that statistically the stationary configurations before and after
the sudden bias change are indistingishable.
As a consequence, no slow relaxation processes follow the change of hopping 
probabilities.
As we have confirmed with our Monte Carlo simulations, any macroscopic 
observables such as the mean particle current assume their new stationary 
values essentially instantaneously after the bias reset.

\subsection{Strongly correlated initial conditions}

\begin{figure}[t]
  \includegraphics[width=\columnwidth,clip]{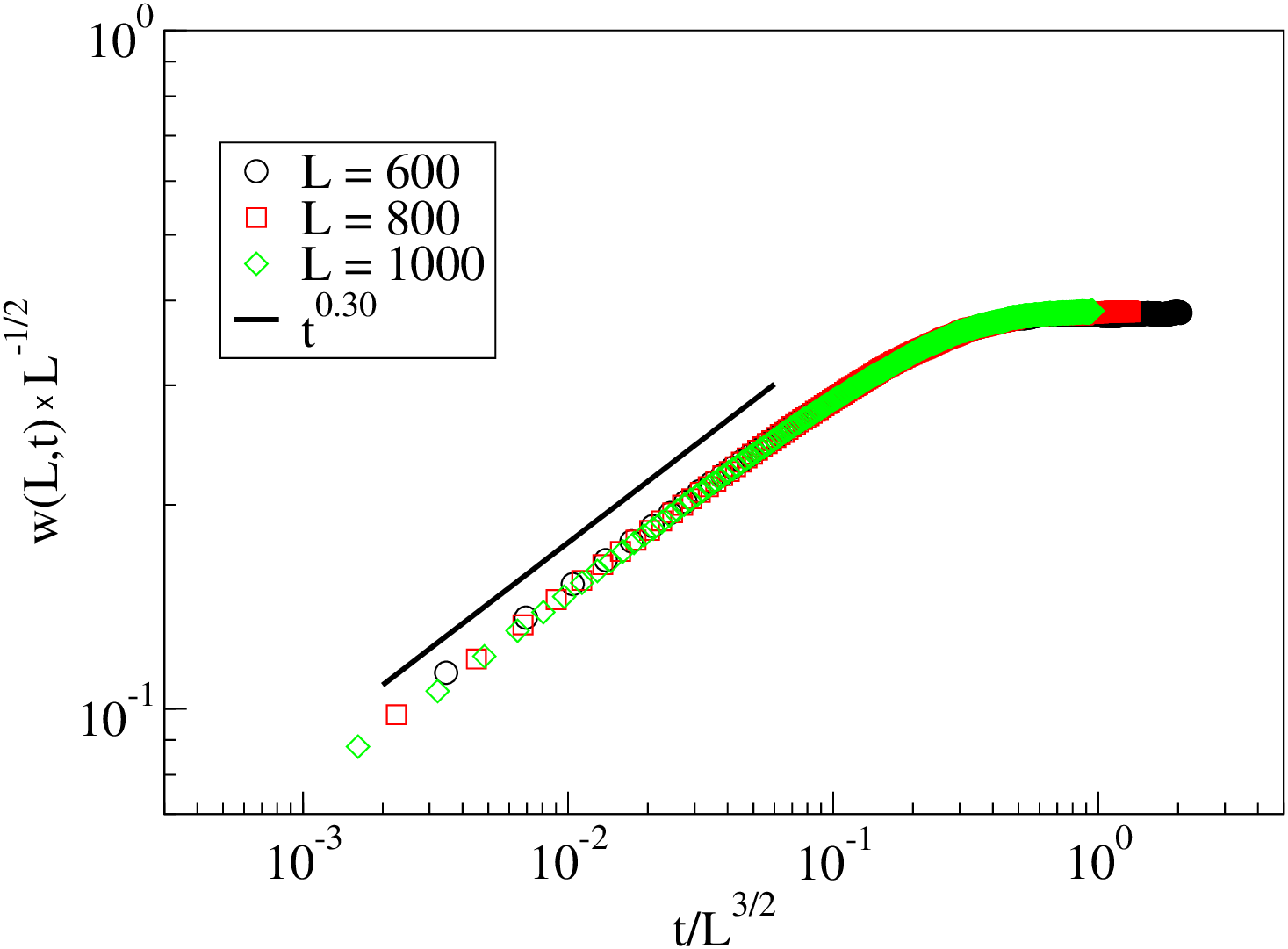}
  \caption{({\it Color online.}) Finite-size scaling for the interface width of
  the one-dimensional growth model that corresponds to the TASEP, for system
  sizes $L = 600$, $800$, and $1000$, see Eq.~(\ref{wfsc}).  
  The data are averaged over 50,000 realizations each. \label{widthfss}}
\end{figure}

In order to construct initial conditions that generate non-stationary 
relaxation with broken time translation invariance, we consider the mapping of
the TASEP to a surface growth problem in the KPZ universality class 
\cite{meak}.
In the growth model, one obtains slow relaxation towards the stationary state 
if the process is initiated with a flat surface $h(x,0) = 0$ everywhere, and 
roughening subsequently commences on small length scales 
$\sim t^{1/z_\parallel}$.
In the TASEP picture, this initial state is accomplished at half-filling 
$\rho = 1/2$ by placing the particles alternatingly on every other site. 
The mean-square interface width of the corresponding finite one-dimensional 
growth model can be obtained from our TASEP Monte Carlo data on a ring of
length $L$ via Eq.~(\ref{mapp}),
\begin{equation}
  w(L,t)^2 = \Big\langle h(x,t)^2 \Big\rangle_L = \frac{1}{L} \displaystyle 
  \sum_{x=1}^L \Bigl[ x - 2 \displaystyle \sum_{y=1}^x n(y,t) \Bigr]^2 ,
\end{equation}
where $w(L,0)^2 = 0$.
The standard finite-size scaling relation for the interface width reads 
\cite{family}
\begin{equation}
  \label{wfsc}
  w(L,t) = L^\alpha W(t / L^{z_\parallel}) ,
\end{equation}
with $\alpha = 1/2$ and $z_\parallel = 3/2$ in the KPZ universality class, 
which implies the initial growth law 
$w(L,t) \sim t^{\alpha / z_\parallel} = t^{1/3}$, valid up to times
$t \approx L^{z_\parallel}$ \cite{kpz}.
Our Monte Carlo simulation results nicely confirm the scaling relation 
(\ref{wfsc}), as evidenced in Fig.~\ref{widthfss} by the satisfactory data 
collapse for various system sizes with the expected scaling exponents (compare
also the analogous mode-coupling results in Fig.~3 of Ref.~\cite{uwehwa}, and 
the simulation data shown in Fig.~1 of Ref.~\cite{Queiroz}).
Our measurements yield an initial growth exponent $0.30$, slightly different 
from the expected value $1/3$, likely caused by sizeable corrections to scaling
in our data.  
Our finite size scaling plot of the width matches the simulation data of 
Ref.~\cite{Queiroz}.

Starting from the highly correlated initial configuration of alternatingly
occupied sites in the TASEP that corresponds to a flat surface in the growth
model representation, time translation invariance is broken during the slow 
approach to the stationary regime.
The density-density or height-height correlation functions then depend 
explicitly on both time arguments $t$ and $t'$.
Henceforth we shall refer to the second argument as waiting time $s$, and 
consider the short-time scaling limit $s \ll t$ \cite{pleimbook}.
By means of a careful renormalization group analysis, Krech \cite{krech} 
established the following scaling form for the two-time height-height 
correlation function in this limit:
\begin{equation}
  C_{\rm IS}(x,t,s \ll t) = \left( \frac{s}{t} \right)^\theta |x|^{2\alpha} 
  F(t/x^{z_\parallel}) ,
\end{equation}
which generalizes Eq.~(\ref{esc}) that is valid in the stationary regime.
Here, the initial slip exponent $\theta = (d+4) / z_\parallel - 2$ can be
expressed in terms of the standard dynamic scaling exponent $z$.
This follows since the initial-time surface does not induce any novel
singularities, as a consequence of the momentum dependence of the nonlinear
vertex in the KPZ problem, and thus no additional renormalizations are 
required.
\begin{figure}[t]
  \includegraphics[width=\columnwidth,clip]{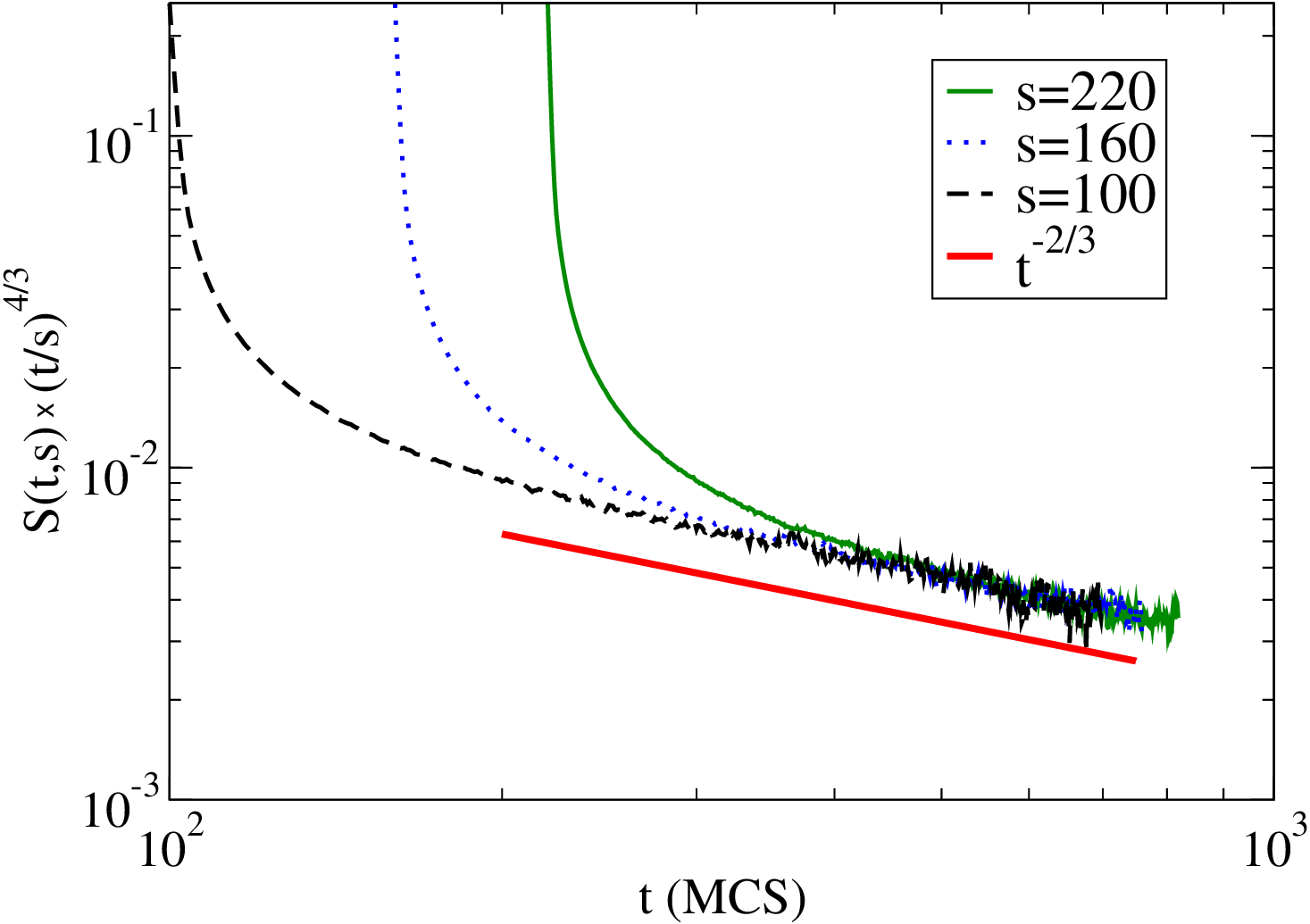}
  \caption{({\it Color online.}) Scaling plot for the density auto-correlation 
  function for a one-dimensional driven lattice gas (TASEP) with length 
  $L = 1000$ and density $\rho = 1/2$ in the non-equilibrium relaxation regime,
  see Eq.~(\ref{isdd}), measured for  waiting times $s = 100$, $160$, and 
  $220$.   
  The data are averaged over 60,000 realizations. \label{isfig}}
\end{figure}

In one dimension, the renormalization group predicts $\theta = 4/3$.
For the TASEP density-density correlation function, we obtain by means of
Eq.~(\ref{hhddrelation})
\begin{eqnarray}
  S_{\rm IS}(x,t,s \ll t) &= \left( \displaystyle \frac{s}{t} \right)^\theta 
  |x|^{2\alpha-2} f(t/x^{z_\parallel}) \nonumber \\
  &= \left( \displaystyle \frac{s}{t}\right)^{4/3} |x|^{-1} f(t/x^{3/2}) .
\end{eqnarray}
In order to arrive at the temporal behavior of the auto-correlation function,
we must require that the scaling function $f(\xi) \sim \xi^{-2/3}$ as its 
argument $\xi \to \infty$, whence
\begin{equation}
  \label{isdd}
  S_{\rm IS}(0,t,s \ll t) = \left( \frac{s}{t} \right)^{4/3} t^{-2/3} .
\end{equation}
Our Monte Carlo auto-correlation data displayed in Fig.~\ref{isfig}, obtained 
for various waiting times $s$, convincingly confirm the scaling predicted by 
Eq.~(\ref{isdd}) for $s \ll t$.
\begin{figure}[t]
  \includegraphics[width=\columnwidth,clip]{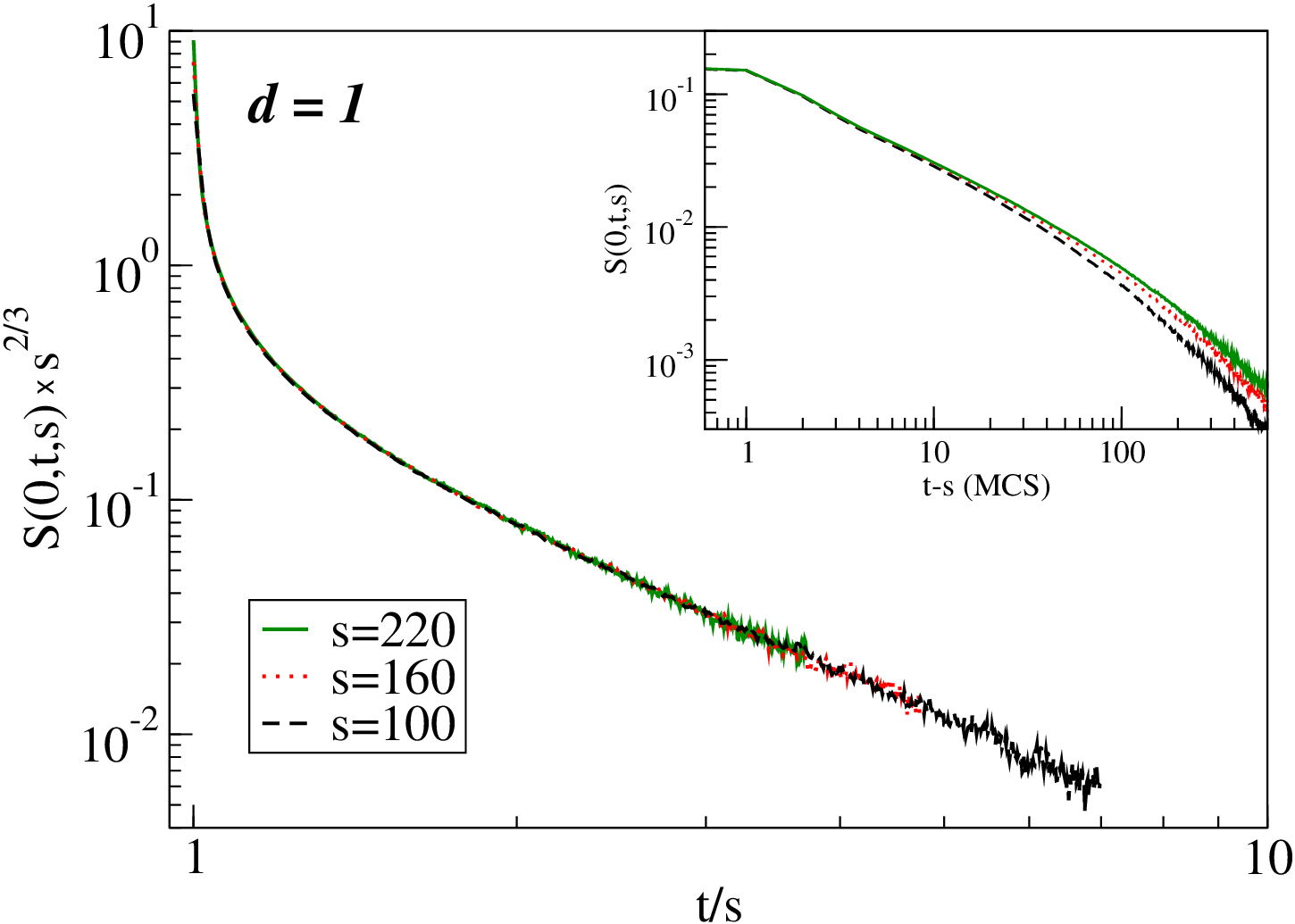}
  \caption{({\it Color online.}) Simple aging scaling plot, see 
  Eq.~(\ref{agescaledds}), for the density-density auto-correlation function 
  for a one-dimensional driven lattice gas (TASEP) of length $L = 1000$, at 
  density $\rho = 1/2$, obtained from Monte Carlo simulation data with 
  different waiting times $s = 100$, $160$, and $220$. 
  The data sets are averaged over 60,000 realizations each. \label{1dage}}
\end{figure}

\begin{figure}[t]
  \includegraphics[width=\columnwidth,clip]{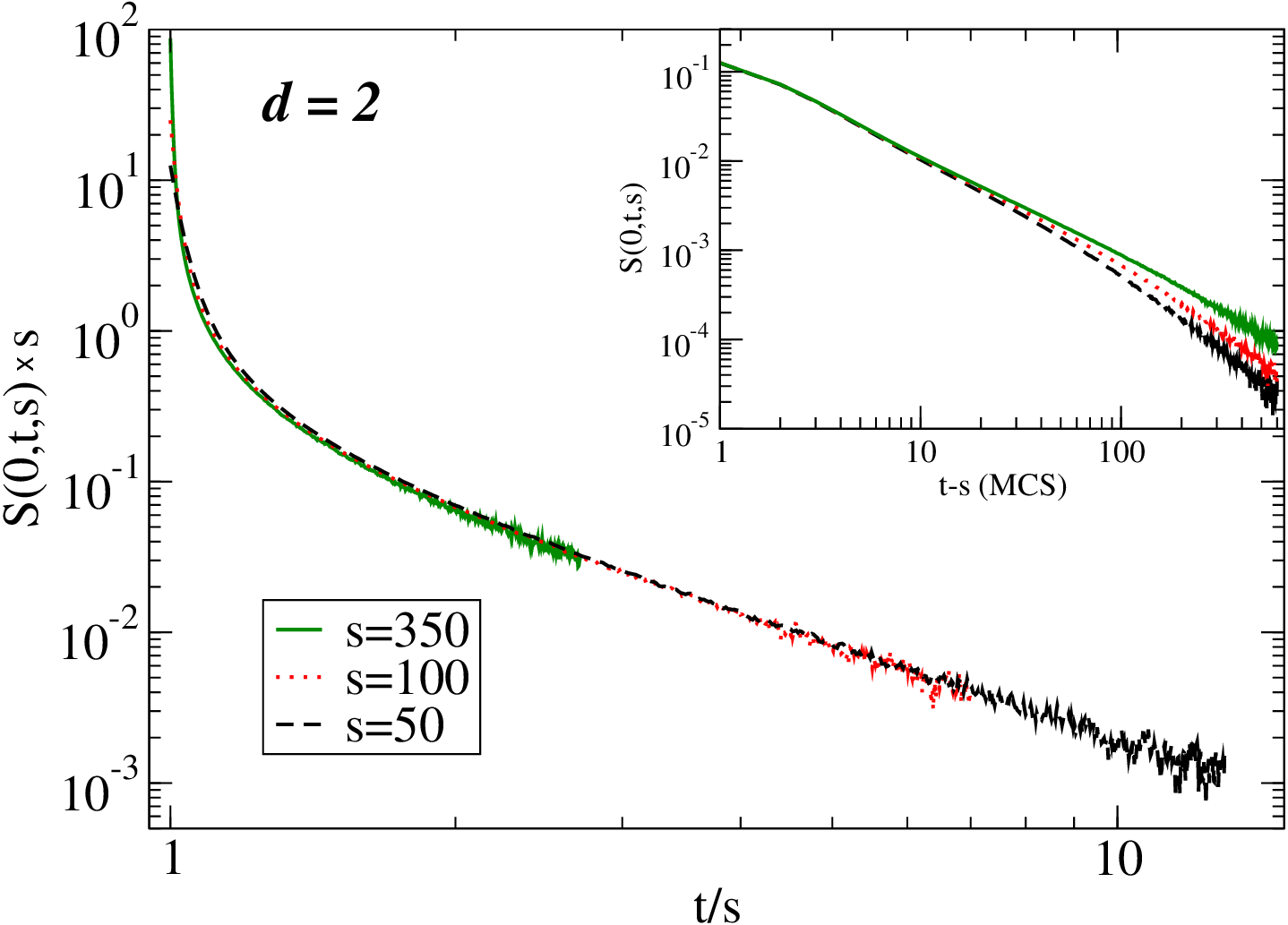}
  \caption{({\it Color online.}) Density auto-correlation function for a 
  two-dimensional driven lattice gas of size $L_x = L_y = 128$ for density 
  $\rho = 1/2$ in the aging scaling regime for waiting times $s = 50$, $100$,
  and $350$.  
  Each data set is averaged over 160,000 realizations. \label{2dage}}
\end{figure}

\begin{figure}[t]
  \includegraphics[width=\columnwidth,clip]{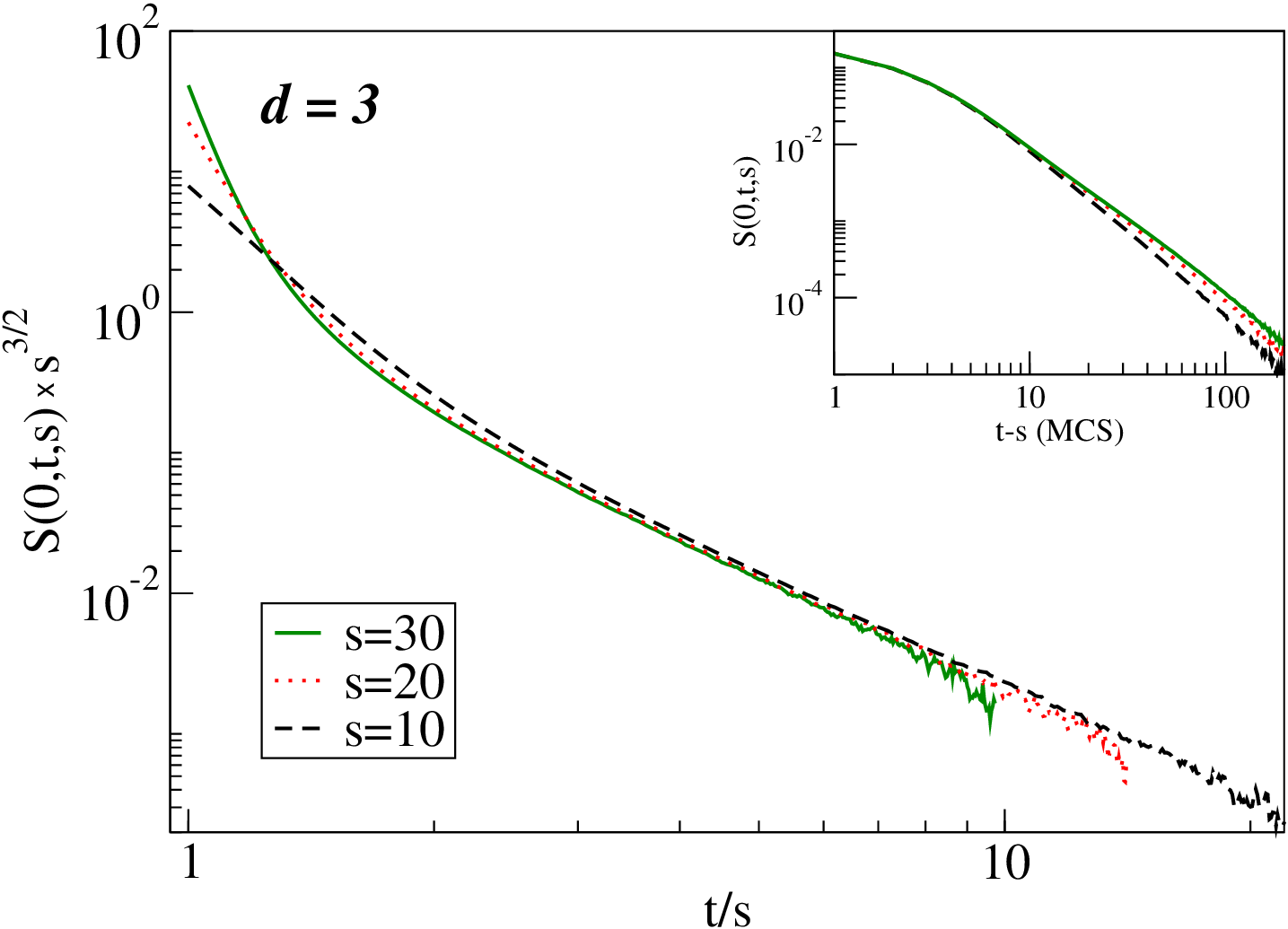}
  \caption{({\it Color online.}) Density auto-correlation function for a 
  three-dimensional driven lattice gas of size $L_x = L_y = L_z = 20$ for 
  density $\rho=1/2$ in the aging scaling regime for waiting times $s = 10$, 
  $20$, and $30$.  
  Each data set is averaged over 4,200,000 realizations. \label{3dage}}
\end{figure}

Finally, we explore the scaling properties of the two-time height and density
auto-correlation functions in the physical aging regime.
As shown for the KPZ growth model, no additional renormalizations, and hence no
independent new scaling exponents are required in the non-stationary relaxation
regime in one dimension \cite{krech}.
Since this is essentially a consequence of the underlying conserved dynamics,
we surmise that we can generalize the scaling laws (\ref{scalehh}) and 
(\ref{xtgen})  to
\begin{eqnarray}
  \label{csc}
  C(x,t,s) = b^{-2 \alpha} C(bx,b^{z_\parallel} t, b^{z_\parallel} s) , \\
  S(x,t,s) = b^{d + \Delta} S(b {\vec x}_\perp, b^{1+\Delta} x_\parallel, 
  b^z t, b^z s) , 
\end{eqnarray}
with $z = z_\parallel (1 + \Delta)$ and 
$2 z (1 - \alpha) = z_\parallel (d + \Delta)$.
Letting $x \to 0$ and setting $b = s^{-1/z_\parallel}$, Eq.~(\ref{csc}) yields
for the growth model height-height auto-correlation function
\begin{equation}
  C(0,t,s)= s^{2 \alpha / z_\parallel} f_C(t/s) = s^{2/3} f_C(t/s) .
\label{agescale}
\end{equation}
Recent Monte Carlo simulations have confirmed this simple aging scaling 
scenario \cite{pleimbook} for the KPZ universality class 
\cite{krug,michelnew,A4,yenglobal}.
In a similar vein, for the density-density auto-correlation function in the 
driven lattice gas, via matching $b = s^{-1/z}$ we obtain 
Eq.~(\ref{agescaledds}), with the scaling exponent $\zeta$ given in 
Eq.~(\ref{zeta}).
Figure~\ref{1dage} confirms the simple aging scenario according to 
Eq.~(\ref{agescaledds}) through the nice data collapse of our Monte Carlo
simulation results for various waiting times in one dimension, where 
$\zeta = 2/3$.
We have also performed driven lattice gas simulations in two and three 
dimensions with alternating particle initial conditions on periodic lattices.
As demonstrated in Figs.~\ref{2dage} and \ref{3dage}, we obtain simple-scaling
data collapse with the expected scaling exponent $\zeta = 1$ for $d = 2$ and
$\zeta = 3/2$ for $d = 3$, respectively.

\section{Conclusion}

We have performed extensive precise Monte Carlo simulations of driven lattice 
gases with exclusion on periodic lattices, and studied in detail the behavior 
of the density correlations in one, two, and three dimensions.

For the one-dimensional case, we have investigated interesting recurrence
oscillations away from half-filling, and explained the minimum value in the 
density auto-correlation function observed for small systems.
We also studied how varying the hopping bias in the ASEP affects the crossover 
to the long-time power law decay of the density auto-correlation function.
While our large-scale simulations confirm standard finite-size scaling 
behavior, the effective exponent that describes the temporal decay of the
auto-correlations approaches the asymptotic value exceedingly slowly, even for 
the TASEP, and displays very unusual finite-size corrections to scaling.
Our Monte Carlo simulation results confirm all expected non-trivial scaling 
exponents for the driven lattice gas, in accord with the scaling properties of
equivalent growth models in the KPZ universality class.
This includes predictions for the simple aging scaling scenario during the 
non-stationary relaxation from strongly correlated initial conditions.
Our data thus support the assertion that in driven lattice gas with exclusion, 
the entire universal scaling regime is governed by a single non-trivial 
exponent $\Delta$ (or equivalently, $z_\parallel$ or $\zeta$).
Specifically, no novel scaling exponent is required to capture non-equilibrium 
aging properties in this system.
We have moreover numerically determined several different universal scaling 
functions.

We extended our Monte Carlo simulations to two and three dimensions. 
Aside from logarithmic corrections at the critical dimension $d_c = 2$, we have
confirmed mean-field scaling behavior for the density auto-correlation function
in both the stationary long-time and the non-stationary aging regimes.

\begin{acknowledgments}
This work was in part supported by the U.S. Department of Energy, Office of 
Basic Energy Sciences (DOE--BES) under grant no. DE-FG02-09ER46613.
We would like to thank Mustansir Barma, Yen-Liang Chou, Ulrich Dobramysl, 
Shivakumar Jolad, Thierry Platini, Michel Pleimling, Beate Schmittmann, 
Matthew Shimer, and Royce Zia for very helpful discussions and insightful 
suggestions.
\end{acknowledgments}

\bibliographystyle{h-physrev}

\begin{thebibliography}{10}

\bibitem{beatebook}
B.~Schmittmann and R.~K.~P. Zia,
\newblock {\em Statistical Mechanics of Driven Diffusive Systems edited by C.
  Domb and J. L. Lebowitz} (Academic Press, London, 1995).

\bibitem{hh_rev}
T.~Halpin-Healy and Y.-C. Zhang,
\newblock Phys. Rep. {\bf 254}, 215  (1995).

\bibitem{derr1}
B.~Derrida,
\newblock Phys. Rep. {\bf 301}, 65  (1998).

\bibitem{guntbook}
G.~M. Sch{\"{u}}tz,
\newblock {\em Phase Transitions and Critical Phenomena Volume 19 edited by C.
  Domb and J. L. Lebowitz} (Academic Press, London, 2000).

\bibitem{kirone_rev}
K.~Mallick,
\newblock J. Stat. Mech. , P01024 (2011).

\bibitem{dhar}
D.~Dhar,
\newblock {Phase Transitions} {\bf {9}}, {51} ({1987}).

\bibitem{bethe_1}
L.-H. Gwa and H.~Spohn,
\newblock Phys. Rev. Lett. {\bf 68}, 725 (1992).

\bibitem{bethe_2}
L.-H. Gwa and H.~Spohn,
\newblock Phys. Rev. A {\bf 46}, 844 (1992).

\bibitem{spec_osc3}
O.~Golinelli and K.~Mallick,
\newblock J. Phys. A: Math. Gen. {\bf 37}, 3321 (2004).

\bibitem{spec_osc2}
D.~Kim,
\newblock Phys. Rev. E {\bf 52}, 3512 (1995).

\bibitem{spec_osc1}
O.~Golinelli and K.~Mallick,
\newblock J. Phys. A: Math. Gen. {\bf 38}, 1419 (2005).

\bibitem{exact1}
G.~Sch{\"{u}}tz,
\newblock J. Stat. Phys. {\bf 88}, 427 (1997).

\bibitem{exact2}
V.~B. Priezzhev,
\newblock Phys. Rev. Lett. {\bf 91}, 050601 (2003).

\bibitem{Forster}
D.~Forster, D.~R. Nelson, and M.~J. Stephen,
\newblock Phys. Rev. A {\bf 16}, 732 (1977).

\bibitem{kpz}
M.~Kardar, G.~Parisi, and Y.-C. Zhang,
\newblock Phys. Rev. Lett. {\bf 56}, 889 (1986).

\bibitem{Jan}
H.~K. Janssen and B.~Schmittmann,
\newblock Z. Phys. B {\bf 63}, 517 (1986).

\bibitem{mc_4}
H.~van Beijeren, R.~Kutner, and H.~Spohn,
\newblock Phys. Rev. Lett. {\bf 54}, 2026 (1985).

\bibitem{mc_2}
T.~Hwa and E.~Frey,
\newblock Phys. Rev. A {\bf 44}, R7873 (1991).

\bibitem{uwehwa}
E.~Frey, U.~C. T{\"{a}}uber, and T.~Hwa,
\newblock Phys. Rev. E {\bf 53}, 4424 (1996).

\bibitem{mc_5}
F.~Colaiori and M.~A. Moore,
\newblock Phys. Rev. Lett. {\bf 86}, 3946 (2001).

\bibitem{cola2}
F.~Colaiori and M.~A. Moore,
\newblock Phys. Rev. E {\bf 63}, 057103 (2001).

\bibitem{mc_1}
B.~Hu and G.~Tang,
\newblock Phys. Rev. E {\bf 66}, 026105 (2002).

\bibitem{mc_3}
L.~Canet and M.~A. Moore,
\newblock Phys. Rev. Lett. {\bf 98}, 200602 (2007).

\bibitem{cola}
F.~Colaiori and M.~A. Moore,
\newblock Phys. Rev. E {\bf 65}, 017105 (2001).

\bibitem{spohnlong}
M.~Pr{\"{a}}hofer and H.~Spohn,
\newblock J. Stat. Phys. {\bf 115}, 255 (2004).

\bibitem{schwartz_1}
E.~Katzav and M.~Schwartz,
\newblock Phys. Rev. E {\bf 69}, 052603 (2004).

\bibitem{rg_2}
J.~Krug, P.~Meakin, and T.~Halpin-Healy,
\newblock Phys. Rev. A {\bf 45}, 638 (1992).

\bibitem{rg_1}
T.~Nattermann and L.-H. Tang,
\newblock Phys. Rev. A {\bf 45}, 7156 (1992).

\bibitem{fog_1}
H.~C. Fogedby,
\newblock J. Phys.: Condens. Matter {\bf 14}, 1557 (2002).

\bibitem{scaling_exact_1}
P.~L. Ferrari and H.~Spohn,
\newblock Commun. Math. Phys. {\bf 265}, 1 (2006).

\bibitem{krech}
M.~Krech,
\newblock Phys. Rev. E {\bf 55}, 668 (1997).

\bibitem{pleimbook}
M.~Henkel and M.~Pleimling,
\newblock {\em Ageing and Dynamical Scaling Far from Equilibrium},
  Nonequilibrium phase transitions Vol.~2 (Springer, Dordrecht, 2010).

\bibitem{krug}
H.~Kallabis and J.~Krug,
\newblock Europhys. Lett. {\bf 45}, 20 (1999).

\bibitem{A4}
S.~Bustingorry,
\newblock J. Stat. Mech. , P10002 (2007).

\bibitem{michelnew}
M.~Henkel, J.~D. Noh, and M.~Pleimling,
\newblock Unpublished  (2010).

\bibitem{yenglobal}
Y.-L. Chou and M.~Pleimling,
\newblock J. Stat. Mech , P08007 (2010).

\bibitem{A1}
A.~R{\"{o}}thlein, F.~Baumann, and M.~Pleimling,
\newblock Phys. Rev. E {\bf 74}, 061604 (2006).

\bibitem{A3}
S.~Bustingorry, L.~F. Cugliandolo, and J.~L. Iguain,
\newblock J. Stat. Mech., P09008 (2007).

\bibitem{A2}
Y.-L. Chou, M.~Pleimling, and R.~K.~P. Zia,
\newblock Phys. Rev. E {\bf 80}, 061602 (2009).

\bibitem{sas_lett}
T.~Sasamoto,
\newblock J. Phys. A: Math. Gen. {\bf 38}, L549 (2005).

\bibitem{sas_fluc}
A.~Borodin, P.~Ferrari, M.~Pr{\"{a}}hofer, and T.~Sasamoto,
\newblock J. Stat. Phys. {\bf 129}, 1055 (2007).

\bibitem{time_dep}
A.~M. Povolotsky and V.~B. Priezzhev,
\newblock J. Stat. Mech. , P08018 (2007).

\bibitem{sas_pape}
T.~Sasamoto,
\newblock Eur. Phys. J. B {\bf 64}, 373 (2008).

\bibitem{New}
M.~E. Newman and G.~T. Barkema,
\newblock {\em Monte Carlo Methods in Statistical Physics} (Oxford University
  Press, Oxford, 1999).

\bibitem{Borz}
A.~B. Bortz, M.~H. Kalos, and J.~L. Lebowitz,
\newblock J. Comput. Phys. {\bf 17}, 10  (1975).

\bibitem{meak}
P.~Meakin, P.~Ramanlal, L.~M. Sander, and R.~C. Ball,
\newblock Phys. Rev. A {\bf 34}, 5091 (1986).

\bibitem{spohn_proc}
M.~Pr{\"{a}}hofer and H.~Spohn,
\newblock Current fluctuations for the totally asymmetric simple exclusion
  process,
\newblock in {\em In and Out of Equilibrium}, Progr. Probab. No. ~51, pp.
  185--204, Birkh{\"{a}}user, Boston, MA, 2002.

\bibitem{adams}
D.~A. Adams, R.~K.~P. Zia, and B.~Schmittmann,
\newblock Phys. Rev. Lett. {\bf 99}, 020601 (2007).

\bibitem{cook}
L.~J. Cook and R.~K.~P. Zia,
\newblock J. Stat. Mech. , P07014 (2010).

\bibitem{oscillate}
S.~Gupta, S.~N. Majumdar, C.~Godr\`eche, and M.~Barma,
\newblock Phys. Rev. E {\bf 76}, 021112 (2007).

\bibitem{exact3}
M.~Makhanova and V.~Priezzhev,
\newblock Theoret. and Math. Phys. {\bf 146}, 421 (2006).

\bibitem{Pier}
P.~Pierobon, A.~Parmeggiani, F.~von Oppen, and E.~Frey,
\newblock Phys. Rev. E {\bf 72}, 036123 (2005).

\bibitem{Juh}
R.~Juh{$\acute{a}$}sz and L.~Santen,
\newblock J. Phys. A: Math. Gen. {\bf 37}, 3933 (2004).

\bibitem{monte_1}
L.~H. Tang,
\newblock J. Stat. Phys. {\bf 67}, 819 (1992).

\bibitem{MBarma}
M.~Barma, M.~D. Grynberg, and R.~B. Stinchcombe,
\newblock J. Phys-Condens. Mat. {\bf 19}, 065112 (2007).

\bibitem{family}
F.~Family and T.~Vicsek,
\newblock J. Phys. A: Math. Gen. {\bf 18} (1985).

\bibitem{Queiroz}
S.~L.~A. de~Queiroz and R.~B. Stinchcombe,
\newblock Phys. Rev. E {\bf 78}, 031106 (2008).

\end{thebibliography}

\end{document}